\documentclass[sigconf]{acmart}
\usepackage[linesnumbered,ruled,vlined]{algorithm2e}
\usepackage[noend]{algpseudocode}
\usepackage{amsmath}
\usepackage{stmaryrd}
\usepackage{url}
\usepackage{listings}
\usepackage{mdwlist}
\usepackage{pifont}
\usepackage{graphicx}
\usepackage{wrapfig}
\usepackage{fancyvrb}
\usepackage{paralist}
\usepackage{caption}
\usepackage{comment}
\usepackage{bbm}
\usepackage{dsfont}
\usepackage{mathrsfs}
\usepackage[T1]{fontenc}
\usepackage{esvect}
\usepackage{array}
\usepackage{multirow}
\usepackage{rotating}
\usepackage{makecell}
\usepackage{hyperref}
\usepackage{amsfonts}
\usepackage{xparse}
\usepackage[export]{adjustbox}
\usepackage{tabularx}
\usepackage{multicol}
\usepackage{acronym}
\usepackage{subcaption}
\usepackage[]{soul}
\usepackage[]{todonotes}
\usepackage{multicol}
\usepackage{float}
\usepackage[framemethod=tikz]{mdframed}
\usepackage{mathpartir}
\usepackage{xfp}
\usepackage{xintfrac}
\usepackage{amsthm}
\usepackage{mfirstuc}
\usepackage{booktabs}
\usepackage{colortbl}
\usepackage{stackengine}
\usepackage{xspace}
\usepackage{afterpage}

\usepackage[nomessages]{fp}
\usepackage[autolanguage]{numprint}
\usepackage{fmtcount}

\usepackage[subtle]{tex/savetrees}

\definecolor{hgreen}{RGB}{160,255,218}
\definecolor{hred}{RGB}{255,204,204}
\definecolor{babypink}{rgb}{0.96, 0.76, 0.76}
\definecolor{bulgarianrose}{rgb}{0.59, 0.0, 0.09}

\mdfdefinestyle{graybox}{
	roundcorner=1mm,%
	backgroundcolor=gray!30
}

\newtheoremstyle{bfnote}%
{}{}
{}{}
{\bfseries}{}
{ }{\thmname{#1}\thmnumber{ #2}\thmnote{ (#3)}}
\theoremstyle{bfnote}

\graphicspath{{figs/}}

\tikzset{inlinenotestyle/.append style={align=justify}}
\SetNoFillComment
\DontPrintSemicolon

\SetCommentSty{mycommfont}
\SetFuncSty{texttt}

\newcolumntype{C}[1]{>{\centering\arraybackslash}m{#1}}
\newcolumntype{L}[1]{>{\raggedleft\arraybackslash}m{#1}}
\newcolumntype{R}[1]{>{\raggedright\arraybackslash}m{#1}}

\newenvironment{conditions*}
{\par\vspace{\abovedisplayskip}\noindent
	\tabularx{\columnwidth}{>{$}l<{$} @{${}={}$} >{\raggedright\arraybackslash}X}}
{\endtabularx\par\vspace{\belowdisplayskip}}

\usepackage{makecell}
\usepackage{multirow}
\newcommand{\etal}{\textit{et al.}}

\newcommand{\etc}{\textit{etc.}}
\newcommand{\eg}{\textit{e.g.}}
\newcommand{\viz}{\textit{viz.}}
\newcommand{\ie}{\textit{i.e.}}

\newcommand{\tick}{{\color{deepgreen}\ding{51}}}
\newcommand{\cross}{{\color{red}\ding{55}}}

\let\num\numprint

\newcommand{\Tbl}[1]{\ensuremath{\sf Table~\ref{#1}}}
\newcommand{\Fig}[1]{\ensuremath{\sf Figure~\ref{#1}}}
\newcommand{\Sec}[1]{\ensuremath{\sf Section~\ref{#1}}}
\newcommand{\Adx}[1]{\ensuremath{\sf Appendix~\ref{#1}}}

\NewDocumentCommand{\lstp}{m  m O{}}{\ensuremath{\sf Listings~\ref{#1} ~\text{and} ~\ref{#2}}}
\NewDocumentCommand{\Lines}{m  m O{}}{\ensuremath{\sf Line~ #1 ~\text{and} ~#2}}

\stackMath

\newcounter{sqindex}

\definecolor{darkorchid}{rgb}{0.6, 0.2, 0.8}
\definecolor{boxgreen}{HTML}{38761d}
\definecolor{boxred}{HTML}{ff0000}
\definecolor{instrblue}{HTML}{cfe2f3}
\definecolor{instrgreen}{HTML}{b6d7a8}
\definecolor{edgered}{HTML}{ff0000}
\definecolor{bgcode}{gray}{0.95}
\definecolor{deepgreen}{HTML}{259D04}

\newcommand{\defnote}[3]{
	\newcounter{#1}
	
	\expandafter
	\newcommand\csname cmt#1\endcsname[1]{
		\refstepcounter{#1} 
		\textcolor{#3}{\textbf{#2 [\csname the#1\endcsname]:}} %
		\hl{\textbf{##1.}}} %
	
	\expandafter
	\newcommand\csname cmtdel#1\endcsname[2]{
		\refstepcounter{#1} 
		\textcolor{#3}{\textbf{#2 [\csname the#1\endcsname]:} ##2} %
		\st{\textbf{##1}}} %

	\expandafter
	\newcommand\csname #1\endcsname[1]{%
		\refstepcounter{#1}%
		{%
			\todo[color={#3!30},inline]{%
				\textbf{#2 [\csname the#1\endcsname]:}~##1}%
	}}
}

\def\numwithsuffix#1{%
\xintifGt{#1}{1000000000}{\fpeval{round(#1/1000000000,1)}B}
{\xintifGt{#1}{1000000}{\fpeval{round(#1/1000000,1)}M}
{\xintifGt{#1}{1000}{\fpeval{round(#1/1000,1)}K}
{#1}}}}

\DeclareRobustCommand\boxlabeled[1]{\tikz[baseline=(char.base)]\node[draw,rectangle,fill=green,inner sep=1.5pt](char) {\textcolor{black}{#1}};}

\newcommand{\irule}[2]%
   {\mkern-2mu\displaystyle\frac{#1}{\vphantom{,}#2}\mkern-2mu}

\makeatletter
\def\endthebibliography{%
	\def\@noitemerr{\@latex@warning{Empty `thebibliography' environment}}%
	\endlist
}
\makeatother

\newcommand{\reentrancy}{{reentrancy}\xspace}
\newcommand{\tod}{{transaction order dependence}\xspace}
\newcommand{\ether}{{Ether}}
\newcommand{\bc}{{blockchain}}

\newcommand{\smart}{smart contract}
\newcommand{\sender}{\texttt{msg.sender}}
\newcommand{\ethereum}{Ethereum\xspace}
\newcommand{\polygon}{Polygon\xspace}
\newcommand{\etherscan}{{\sc Etherscan}\xspace}
\newcommand{\coinbase}{{\sc Coinbase}\xspace}
\newcommand{\binance}{{\sc Binance}\xspace}
\newcommand{\opensea}{{\sc OpenSea}\xspace}
\newcommand{\sorare}{{\sc Sorare}\xspace}
\newcommand{\axie}{{\sc Axie}\xspace}
\newcommand{\foundation}{{\sc Foundation}\xspace}
\newcommand{\nifty}{{\sc Nifty}\xspace}
\newcommand{\valuables}{{\sc Valuables}\xspace}
\newcommand{\decentraland}{{\sc Decentraland}\xspace}
\newcommand{\rarible}{{\sc Rarible}\xspace}
\newcommand{\superrare}{{\sc SuperRare}\xspace}
\newcommand{\cryptokitties}{{\sc CryptoKitties}\xspace}
\newcommand{\cryptopunks}{{\sc CryptoPunks}\xspace}
\newcommand{\godsunchained}{{\sc Gods Unchained}\xspace}
\newcommand{\cryptomotors}{{\sc CryptoMotors}\xspace}

\newcommand{\cryptovoxels}{{\sc CryptoVoxels}\xspace}
\newcommand{\mirandusvaults}{{\sc Mirandus Vaults}\xspace}
\newcommand{\ens}{{\sc Ethereum Name Service}\xspace}

\newcommand{\dappradar}{{\sc DappRadar}\xspace}
\newcommand{\coingecko}{{\sc CoinGecko}\xspace}
\newcommand{\nftartfinance}{{\sc NFT-Art.Finance}\xspace}
\newcommand{\discord}{{\sc Discord}\xspace}
\newcommand{\twitter}{{\sc Twitter}\xspace}

\newcommand{\reddit}{{\sc Reddit}\xspace}

\newcommand{\wyvern}{{\sc Wyvern}\xspace}

\newcommand{\nftm}{NFTM}
\newcommand{\ercnonfungible}{ERC-721}
\newcommand{\ercfungible}{ERC-20}
\newcommand{\dapp}{dApp}
\newcommand{\defii}{DeFi}
\newcommand{\defiposer}{{\sc DefiPoser}\xspace}
\newcommand{\flashboys}{{\sc FlashBoys}\xspace}
\newcommand{\imagehash}{{\sc ImageHash}\xspace}
\newcommand{\github}{{\sc GitHub}\xspace}

\newcommand{\datacollectionstartdate}{June 15, 2021}

\newcommand{\marketplacesbyvolume}{$8$} %
\newcommand{\marketplacesanalyzedfortradingmalpractices}{$7$}
\newcommand{\totalmarketplaces}{$8$}
\newcommand{\totaldappradarmarketplaces}{$35$}

\newcommand{\openseacollections}{236057}

\newcommand{\totalcollections}{238180}

\newcommand{\totalcollectionswithmorethantwoktxvolume}{8869}

\newcommand{\openseatradingvolume}{$4.32$B}
\newcommand{\openseaassets}{12215650}
\newcommand{\openseaexternalassets}{9064767}
\newcommand{\openseaevents}{349911634}

\newcommand{\openseatotalassets}{18250000}

\newcommand{\axietradingvolume}{$1.75$B}
\newcommand{\axieassets}{891238}
\newcommand{\axieevents}{487486}

\newcommand{\cryptopunkstradingvolume}{$1.18$B}
\newcommand{\cryptopunksassets}{9999}
\newcommand{\cryptopunksevents}{172157}

\newcommand{\raribletradingvolume}{$199.42$M}
\newcommand{\raribleassets}{72509}
\newcommand{\raribleevents}{1864997}

\newcommand{\superraretradingvolume}{$106.87$M}
\newcommand{\superrareassets}{28676}
\newcommand{\superrareevents}{198848}

\newcommand{\soraretradingvolume}{$97.42$M}
\newcommand{\sorareassets}{298219}
\newcommand{\sorareevents}{1392292}

\newcommand{\foundationtradingvolume}{$68.19$M}
\newcommand{\foundationassets}{112120}
\newcommand{\foundationevents}{508349}

\newcommand{\niftytradingvolume}{$300.12$M}

\newcommand{\openseaverifiedsellers}{502}
\newcommand{\openseanonverifiedsellers}{124398}
\newcommand{\openseaverifiedcollections}{1805}
\newcommand{\openseanonverifiedcollections}{234112}
\newcommand{\openseaverifiedsellerstotalsales}{114470032}
\newcommand{\openseanonverifiedsellerstotalsales}{2736829199}
\newcommand{\openseaverifiedcollectionstotalsales}{3293911235}
\newcommand{\openseanonverifiedcollectionstotalsales}{403358456}
\newcommand{\openseaverifiedsellerstakendown}{-}
\newcommand{\openseanonverifiedsellerstakendown}{-}
\newcommand{\openseaverifiedcollectionstakendown}{88}
\newcommand{\openseanonverifiedcollectionstakendown}{11182}

\newcommand{\openseaassetswithmetadataurlinallthreecrawls}{3079139}
\newcommand{\openseametadataurlchangedbetweenfirstandsecondcrawl}{89089}
\newcommand{\openseametadataurlchangedbetweensecondandthirdcrawl}{35446}

\newcommand{\totalassetcontracts}{11339}
\newcommand{\totalassetcontractswithsource}{8122}
\newcommand{\totalassetcontractswithoutsource}{3217}
\newcommand{\openseaassetcontractswithsource}{7850}
\newcommand{\openseaassetcontractswithoutsource}{3209}
\newcommand{\openseaassetcontractswithsourcetakendown}{606}
\newcommand{\openseaassetcontractswithoutsourcetakendown}{1765}
\newcommand{\openseaassetcontractswithoutsourcetakendowntradingvolume}{328778895}

\newcommand{\openseatotalauctions}{48862}
\newcommand{\raribletotalauctions}{19109}

\newcommand{\openseatokennottransferredtohighestbidder}{16215}
\newcommand{\raribletokennottransferredtohighestbidder}{15368}

\newcommand{\nonexistentimageurlcachedbyopensea}{2691030}

\newcommand{\escrowcountingenddate}{December 31, 2021}
\newcommand{\foundationescrowednfts}{64079}
\newcommand{\niftyescrowednfts}{90988}
\newcommand{\superrareescrowednfts}{55}

\newcommand{\openseaonchainfeesettlement}{56920}
\newcommand{\superrareonchainfeesettlement}{302}
\newcommand{\raribleonchainfeesettlement}{2777}
\newcommand{\foundationonchainfeesettlement}{5}
\newcommand{\cryptopunksonchainfeesettlement}{814}
\newcommand{\axieonchainfeesettlement}{0}
\newcommand{\sorareonchainfeesettlement}{56}
\newcommand{\royaltypostsalesmodification}{157450}
\newcommand{\royaltypostsalesmodificationincollection}{20802}

\newcommand{\existentipfsimageurl}{907491}
\newcommand{\nonexistentipfsimageurl}{36929}
\newcommand{\existentnonipfsimageurl}{3510828}
\newcommand{\nonexistentnonipfsimageurl}{3908302}
\newcommand{\existentipfsmetadataurl}{557887}
\newcommand{\nonexistentipfsmetadataurl}{55454}
\newcommand{\existentnonipfsmetadataurl}{2082874}
\newcommand{\nonexistentnonipfsmetadataurl}{479429}

\newcommand{\lostnftsalestxvolume}{118294}
\newcommand{\lostnfttradingvolume}{160761805}

\newcommand{\totalcollectionsgreaterthaneightchars}{52399}
\newcommand{\totalsimilarverifiedcollectionpairs}{322}
\newcommand{\similarcollectionnamepairsmanuallyverified}{100}
\newcommand{\similarcollectionnamepairstakendown}{11}
\newcommand{\similarcollectionnamepairsfoundsimilar}{11}
\newcommand{\nftwithidenticalurlsipfs}{356377}
\newcommand{\nftwithidenticalurlsnonipfs}{2082119}
\newcommand{\totalnftimages}{9991013}
\newcommand{\nftwithsimilarimages}{59425}
\newcommand{\truepositivepctsimilarimages}{90}

\newcommand{\washtradecomponentsize}{50}
\newcommand{\washtradesccfrequencythreshold}{10}
\newcommand{\washtradedetected}{9393}
\newcommand{\collectionswashtraded}{5297}
\newcommand{\collectionswashtradedmorethantwoktxvolume}{2569}
\newcommand{\userswashtraded}{17821}
\newcommand{\totalwashtradevolume}{96858093}
\newcommand{\washtradevolumeopensea}{48821949}
\newcommand{\washtradevolumerarible}{47727109}
\newcommand{\washtradevolumesorare}{255805}
\newcommand{\washtradevolumesuperrare}{131}
\newcommand{\washtradevolumecryptopunks}{0}
\newcommand{\washtradevolumeaxie}{0}
\newcommand{\washtradevolumefoundation}{0}
\newcommand{\collectionswithlessthanfivepctwashtrades}{1824}%
\newcommand{\collectionsmorethannintyfivepctwashtrades}{1571}
\newcommand{\totalwashtradevolumebycollectionsmorethannintyfivepctwashtrades}{3407284}
\newcommand{\washtradingmanuallyanalyzed}{100}
\newcommand{\washtradinginstanceswithlessthanequaltotencomponentsize}{9288}

\newcommand{\shillbiddetected}{703}
\newcommand{\collectionsshillbidded}{282}
\newcommand{\usersshillbidded}{1211}
\newcommand{\totalshillbidsellerprofit}{13014662}
\newcommand{\shillbiddingcollectionone}{197}
\newcommand{\shillbiddingcollectionlessthantwenty}{281}
\newcommand{\shillbiddingfoundation}{212}
\newcommand{\shillbiddingcryptovoxels}{11}
\newcommand{\shillbiddingsuperrare}{15}
\newcommand{\shillbiddingmanuallyanalyzed}{100}
\newcommand{\shillbiddingtruepositive}{61}

\newcommand{\bidshieldsdetected}{316}
\newcommand{\collectionsbidshielded}{117}
\newcommand{\usersbidshielded}{471}
\newcommand{\totalshieldedbidamount}{942061}
\newcommand{\bidshieldingminamount}{200.77}
\newcommand{\bidshieldingmaxamount}{152606.31}
\newcommand{\bidshieldingcollectionlessthanten}{113}
\newcommand{\bidshieldamountcryptovoxels}{24519.27}
\newcommand{\bidshieldcountcryptovoxels}{35}
\newcommand{\cryptovoxelsitems}{5800}
\newcommand{\cryptovoxelsvolume}{19200}
\newcommand{\bidshieldcountens}{49}
\newcommand{\bidshieldingmanuallyanalyzed}{100}
\newcommand{\bidshieldingtruepositive}{90}
\newcommand{\bidshieldinginverifiedcollections}{78}

\newcommand{\topsales}{$15$}

\definecolor{verylightgray}{rgb}{.99,.99,.99}
\definecolor{burgundy}{rgb}{0.5, 0.0, 0.13}
\definecolor{brown(web)}{rgb}{0.65, 0.16, 0.16}

\lstdefinelanguage{Solidity}{
	keywords=[1]{anonymous, assembly, assert, balance, break, call, callcode, case, catch, class, constant, continue, constructor, contract, debugger, default, delegatecall, delete, do, else, emit, event, experimental, export, external, false, finally, for, function, gas, if, implements, import, in, indexed, instanceof, interface, internal, is, length, library, log0, log1, log2, log3, log4, memory, modifier, new, payable, pragma, private, protected, public, pure, push, require, return, returns, revert, selfdestruct, send, solidity, storage, struct, suicide, super, switch, then, this, throw, transfer, true, try, typeof, using, value, view, while, with, addmod, ecrecover, keccak256, mulmod, ripemd160, sha256, sha3, , mapping}, %
	keywordstyle=[1]\color{blue},
	keywords=[2]{address, bool, byte, bytes, bytes1, bytes2, bytes3, bytes4, bytes5, bytes6, bytes7, bytes8, bytes9, bytes10, bytes11, bytes12, bytes13, bytes14, bytes15, bytes16, bytes17, bytes18, bytes19, bytes20, bytes21, bytes22, bytes23, bytes24, bytes25, bytes26, bytes27, bytes28, bytes29, bytes30, bytes31, bytes32, enum, int, int8, int16, int24, int32, int40, int48, int56, int64, int72, int80, int88, int96, int104, int112, int120, int128, int136, int144, int152, int160, int168, int176, int184, int192, int200, int208, int216, int224, int232, int240, int248, int256, string, uint, uint8, uint16, uint24, uint32, uint40, uint48, uint56, uint64, uint72, uint80, uint88, uint96, uint104, uint112, uint120, uint128, uint136, uint144, uint152, uint160, uint168, uint176, uint184, uint192, uint200, uint208, uint216, uint224, uint232, uint240, uint248, uint256, var, void, ether, finney, szabo, wei, days, hours, minutes, seconds, weeks, years},	%
	keywordstyle=[2]\color{teal},
	keywords=[3]{block, blockhash, coinbase, difficulty, gaslimit, number, timestamp, msg, data, gas, sig, value, now, tx, gasprice, origin},	%
	keywordstyle=[3]\color{violet} ,
	identifierstyle=\color{black},
	sensitive=false,
	comment=[l]{//},
	morecomment=[s]{/*}{*/},
	commentstyle=\color{brown(web)}\ttfamily,
	stringstyle=\color{red}\ttfamily,
	morestring=[b]',
	morestring=[b]"
}

\setcopyright{acmcopyright}
\acmYear{2022}
\acmDOI{XX.XXXX/XXXXXXX.XXXXXXX}

\acmConference[CCS '22]{ACM Conference on Computer and Communications Security}{November 7-11, 2022}{Los Angeles, USA}

\acmISBN{978-1-4503-XXXX-X/18/06}

\begin{document}
	
	\author{Dipanjan Das}
	\affiliation{%
		\institution{University of California, Santa Barbara}
		\city{Santa Barbara}
		\state{California}
		\country{USA}}
	\email{dipanjan@cs.ucsb.edu}
	
	\author{Priyanka Bose}
	\affiliation{%
		\institution{University of California, Santa Barbara}
		\city{Santa Barbara}
		\state{California}
		\country{USA}}
	\email{priyanka@cs.ucsb.edu}
	
	\author{Nicola Ruaro}
	\affiliation{%
		\institution{University of California, Santa Barbara}
		\city{Santa Barbara}
		\state{California}
		\country{USA}}
	\email{ruaronicola@cs.ucsb.edu}
	
	\author{Christopher Kruegel}
	\affiliation{%
		\institution{University of California, Santa Barbara}
		\city{Santa Barbara}
		\state{California}
		\country{USA}}
	\email{chris@cs.ucsb.edu}
	
	\author{Giovanni Vigna}
	\affiliation{%
		\institution{University of California, Santa Barbara}
		\city{Santa Barbara}
		\state{California}
		\country{USA}}
	\email{vigna@cs.ucsb.edu}
	
	\title{Understanding Security Issues in the NFT Ecosystem}
	
	\begin{abstract}
Non-Fungible Tokens (NFTs) have emerged as a way to collect digital art as well as an investment vehicle.
Despite having been popularized only recently, NFT markets have witnessed several high-profile (and high-value) asset sales and a tremendous growth in trading volumes over the last year.
Unfortunately, these marketplaces have not yet received much security scrutiny.
Instead, most academic research has focused on attacks against decentralized finance (\defii) protocols and automated techniques to detect \smart{} vulnerabilities. 
To the best of our knowledge, we are the first to study the market dynamics and security issues of the multi-billion dollar NFT ecosystem.

In this paper, we first present a systematic overview of how the NFT ecosystem works, and we identify three major actors: marketplaces, external entities, and users.
We then perform an in-depth analysis of the top \totalmarketplaces{} marketplaces (ranked by transaction volume) to discover potential issues, many of which can lead to substantial financial losses.
We also collected a large amount of asset and event data pertaining to the NFTs being traded in the examined marketplaces.
We automatically analyze this data to understand how the entities external to the \bc{} are able to interfere with NFT markets, leading to serious consequences, and quantify the malicious trading behaviors carried out by users under the cloak of anonymity.
\end{abstract}

	
	
	\maketitle
	
	\section{Introduction}
\label{sec:introduction}

A \textit{Non-Fungible Token} (NFT) is an ownership record stored on a \bc{} (such as the \ethereum blockchain).
While digital items, such as pictures and videos, are the most common assets traded as NFTs, the sale of physical assets, \eg, postal stamps~\cite{nft-postage1,nft-postage2}, gold~\cite{nft-gold}, real estate~\cite{nft-real-estate}, physical artwork~\cite{nft-physical-art}, \etc, is also steadily gaining popularity.
In the cryptocurrency world, an NFT is the equivalent of a conventional proof-of-purchase, such as a paper invoice or an electronic receipt.
Among other things, what make NFTs attractive are \textit{verifiability} and \textit{trustless transfer}~\cite{Wang21}.
Verifiability means that sales are recorded as \bc{} transactions, which makes tracking of ownership possible.
In addition, the NFT concept allows for the trading of digital assets between two mutually distrusting parties, as both the crypto payment and the asset transfer happen atomically in a single transaction.

Several NFT marketplaces (\nftm{}s), \eg, \opensea, \rarible, and \axie, emerged in recent years to facilitate buying and selling NFTs. 
This has sparked the interest of both crypto art collectors and traders.
To put things into perspective, \opensea, the largest \nftm, collected $\$236$M USD in platform fees generated out of a trading volume of $\$3.5$B USD~\cite{dune-opensea} in August $2021$ alone.
This is around half of the volume~\cite{opensea-beats-ebay} generated by the e-commerce giant eBay during the same period.
And the all-time combined trading volume of the top three \nftm{}s---\opensea, \axie, and \cryptopunks---surpassed $\$10$B USD in September $2021$~\cite{dappradar}.
Individual NFT sales have also skyrocketed in recent months~\cite{Nadini21}, with nine out of ten of the most expensive sales~\cite{nft-high-value} taking place between February and August $2021$. 
For example, the media widely reported on the digital artist Beeple, who sold an art piece for $\$69.3$M USD; as another example, the first tweet of Twitter CEO Jack Dorsey was sold for $\$2.9$M USD. 
Also, \nftm{}s have surfaced as the most \textit{gas}-eating \ethereum contracts.
For example, \opensea made it to the top of the list of \textit{gas-guzzlers} in \etherscan~\cite{etherscan}, consuming around $20\%$ of the gas spent by the network.

As the NFT space exploded with multi-million dollar sales, cybercriminals and scammers have inevitably flocked to the markets to make quick profits and cheat unsuspecting users.
As a result, numerous NFT scams also made recent headlines. 

\textit{Legitimacy} is one of the big issues with NFTs, as nothing prevents an impostor from ``tokenizing'' and selling someone else's art, while the creator remains oblivious of the fraud.
With the current state of affairs, the onus of verifying the token is on the buyer.
Unfortunately, this is not always easy. 
For instance, in August $2021$, a perpetrator impersonated the popular British graffiti artist Banksy and sold an NFT~\cite{banksy-sales} that featured a ``fake'' art piece by the artist for $\$336$K USD through an online auction.
While \nftm{}s try to thwart such attacks by mandating account validation, typically through an artist's social media presence, another scammer punched a hole through \rarible's verification process and managed to get a fake account associated with the renowned artist Derek Laufman verified~\cite{artist-impersonation}.
Counterfeits NFTs, also called \textit{copycats} or \textit{parody} projects, resemble  reputable collections and purport to have been created by reputable sources. For example, the early NFT project \cryptopunks has numerous clones, such as \textsc{CryptoP\textbf{h}unks}.
In some scenarios, scammers set up unauthorized customer support channels and social media accounts that pretend to be affiliated with \nftm{}s in an effort to steal customer information and compromise accounts~\cite{nft-social-engineering}.
Also, there is evidence of \textit{rug-pulls}, where the owner/creator of an NFT unscrupulously hypes an asset in order to inflate its value, only to cash out, leaving others to suffer from the subsequent decline in value.
One such example is the \textsc{Eternal Beings} collection, which was promoted by the popular American rapper Lil Uzi Vert through his \twitter account with $8.5$M followers.
Soon after the initial investment by the buyers, he deleted all of his tweets, causing the token values to plummet~\cite{lil-uzi-rug-pull}.

With enormous funds flowing into \textit{decentralized finance} (\defii) applications, scams have become lucrative money-making opportunities.
Previous research studied several different aspects of crypto-economic attacks, \eg, financial repercussions due to transaction reordering~\cite{Eskandari20,Daian19,Zhou20,Qin21}, flash loan abuse~\cite{Qin20}, arbitrage opportunities~\cite{Zhou21}, and pump-and-dump schemes~\cite{Kamps18,Gandal18,Xu2019}.
Besides protocol attacks, there also exists a substantial body of work on automated detection of \smart{} vulnerabilities, \eg, \reentrancy, \tod, integer overflows, and unhandled exceptions~\cite{securify,madmax,zeus,mythril,oyente,ethbmc,manticore,contractfuzzer, echidna, sfuzz,sailfish,txspector,ecf}.

To the best of our knowledge, however, the existing literature has not explored the security challenges in the emerging NFT ecosystem, or performed a systematic and comprehensive analysis of the associated threats.
Our work fills that void.
First, we identify three components constituting the NFT ecosystem.
We then analyze each component to discover security, privacy, and usability issues, as well as economic threats.
We hope that our work will be helpful both for NFT marketplaces and their users.
We envision this paper as a guide to help \nftm{}s to avoid mistakes while making users aware of the perils of the NFT space.
In particular, we make the following contributions:

\vspace{0.5mm}
\noindent
\textbf{Anatomy of the NFT ecosystem.}
We systematize the NFT ecosystem, looking at the participating actors ---marketplaces, external entities, and users--- and we analyze their mutual interactions (\Sec{sec:ecosystem}).

\vspace{0.5mm}
\noindent
\textbf{Comprehensive data collection.}
We leverage multiple sources of data, including the \ethereum{} \bc, 
as well as asset and event data sourced from the \nftm{} \dapp{}s, to paint a holistic picture of how the ecosystem operates (\Sec{sec:analysis_approach}). 

\vspace{0.5mm}
\noindent
\textbf{Identifying irregularities in \nftm s.}
We identify flaws in the \nftm{} designs, which, if abused, pose a significant financial risk.

\vspace{0.5mm}
\noindent
\textbf{Identifying issues with external entities.}
We identify the off-chain external entities connected to the NFT ecosystem, and how such entities can pose threats to users (\Sec{sec:external_entity_issues}). 

\vspace{0.5mm}
\noindent
\textbf{Uncovering malicious user behaviors.}
We discover and quantify trading malpractices, such as wash trading, shill bidding, and bid shielding, which are taking place in the top marketplaces.
The insight drawn from our analysis sheds light on some of the prime factors responsible for driving up the recent NFT frenzy (\Sec{sec:user_issues}).

\vspace{0.5mm}
\noindent
\textbf{Releasing code and data.}
We will open-source our analysis framework along with the data we collected to help researchers uncover further interesting insights about the emerging NFT economy.

	\section{Background} 
\label{sec:background}

In this section, we introduce the building blocks of the \ethereum ecosystem, with an emphasis on non-fungible tokens (NFTs) and the economy that has grown around them.

\vspace{0.5mm}
\noindent
\textbf{The \ethereum Blockchain.}
\ethereum is the technology powering the cryptocurrency \ether{} (ETH) and thousands of decentralized applications (\dapp s).
The \ethereum blockchain is a distributed, public ledger where transactions are mined into \textit{blocks} by \textit{miners} who solve cryptographic Proof of Work (PoW) challenges.
In this ecosystem, an \textit{account} is an entity represented by an address that is capable of submitting transactions.
There are two types of accounts in \ethereum: externally owned accounts (EOA), which are controlled by anyone holding the corresponding private key, and contract accounts, which contain executable pieces of code, called \smart{}s.
A \smart{} is a program run by the \ethereum Virtual Machine (EVM), which leverages the blockchain to store its persistent state.
A \textit{transaction} is the transfer of funds between accounts, or an invocation of a contract's public method.
The address that sends the funds or interacts with the contract is denoted by \sender.

\vspace{0.5mm}
\noindent
\textbf{Non-Fungible Token (NFT).}
In the real world, tokens are representations of facts, such as the position in a queue or the authorization to access a facility.
In \ethereum, tokens are digital assets built on top of the blockchain.
Unlike \ether{}, which is the native (built-in) cryptocurrency of the \ethereum blockchain, tokens are implemented by specialized \smart s.
There are two main types of tokens: fungible and non-fungible.
All the copies of a \textit{fungible} token, usually conforming to the \ercfungible{} interface~\cite{erc20}, are identical and interchangeable.
Such tokens can act as a secondary currency within the ecosystem, or can represent someone's stake in an investment. 
On the other hand, all the copies of \textit{non-fungible} tokens, usually conforming to the \ercnonfungible~\cite{erc721} interface, are unique, and each token represents someone's ownership of a specific digital asset, such as ENS domains~\cite{ens_nft} and CryptoKitties~\cite{cryptokitties}, or a physical asset, like a gold bar.

\ercnonfungible~\cite{erc721} is by far the most popular standard for implementing non-fungible tokens on \ethereum.
The standard interface defines a set of mandatory and optional API methods that a token contract needs to implement.
\Fig{fig:erc721} presents a few of those API methods relevant to our discussion.

\begin{figure}[htbp]
	\vspace{-2.5mm}
\begin{lstlisting}[
	backgroundcolor=\color{bgcode},
	caption=,
	numbers=none,
	breakatwhitespace=true,
	language=Solidity]
setApprovalForAll(address _operator, 
                  bool _approved) external
approve (address _approved, 
         uint256 _tokenId) external payable
transferFrom (address _from, address _to, 
              uint256 tokenId) external payable
tokenURI (uint256 _tokenId)
          external view returns (string)
\end{lstlisting}
	\vspace{-4mm}
	\caption{\small Important methods defined in \ercnonfungible{}.}
	\label{fig:erc721}
	\vspace{-4mm}
\end{figure}

Each NFT has its own ID (to keep track of these unique tokens), which is referred to as \texttt{\_tokenId}.
In \ercnonfungible, an \textit{operator} is an entity that can manage all of an NFT owner's assets.
In other words, an NFT owner can delegate the authority to act on her assets to an operator.
Depending on whether the \texttt{\_approved} argument is set, the \texttt{setApprovalForAll()} method either adds or removes the address \texttt{\_operator} from/to the set of the operators authorized by the \sender{} (the NFT's owner).
Unlike an operator, who can operate on all the assets of an owner, \ercnonfungible{} defines a \textit{controller} as an entity who is authorized to operate on one single asset held by an owner.
The \texttt{approve()} method approves the address \texttt{\_approved} as the controller of the asset \texttt{\_tokenId}.
An operator, a controller, or the owner can call the \texttt{transferFrom()} method to transfer the token \texttt{\_tokenId} from the current owner's \texttt{\_from} address to the \texttt{\_to} address.

When an NFT is created (minted), the creator can optionally associate a URL with the NFT. 
That URL, called \texttt{metadata\_url}, should point to a JSON file that conforms to the \ercnonfungible{} Metadata JSON Schema~\cite{erc721}.
The JSON file stores the details of the asset, \eg, its \texttt{name} and \texttt{description}, and also contains an \texttt{image} field storing a URL, called \texttt{image\_url}, that points to the asset.
In this way, an NFT essentially connects an asset with the record of its ownership.
Given a \texttt{\_tokenId}, the associated \texttt{metadata\_url} can be retrieved by querying the \texttt{tokenURI()} API of the contract.
Interestingly, the creation and destruction of NFTs (``minting'' and ``burning'') are not a part of the standard.
Typically, \texttt{mint()} is defined as a public function restricted to the contract creator, and invoked by passing \texttt{metadata\_url} as an argument.
Minting can also be done during contract creation by calling \texttt{mint()} through the contract's constructor.

\begin{figure}[t]
	\centering
	\includegraphics[width=0.5\textwidth]{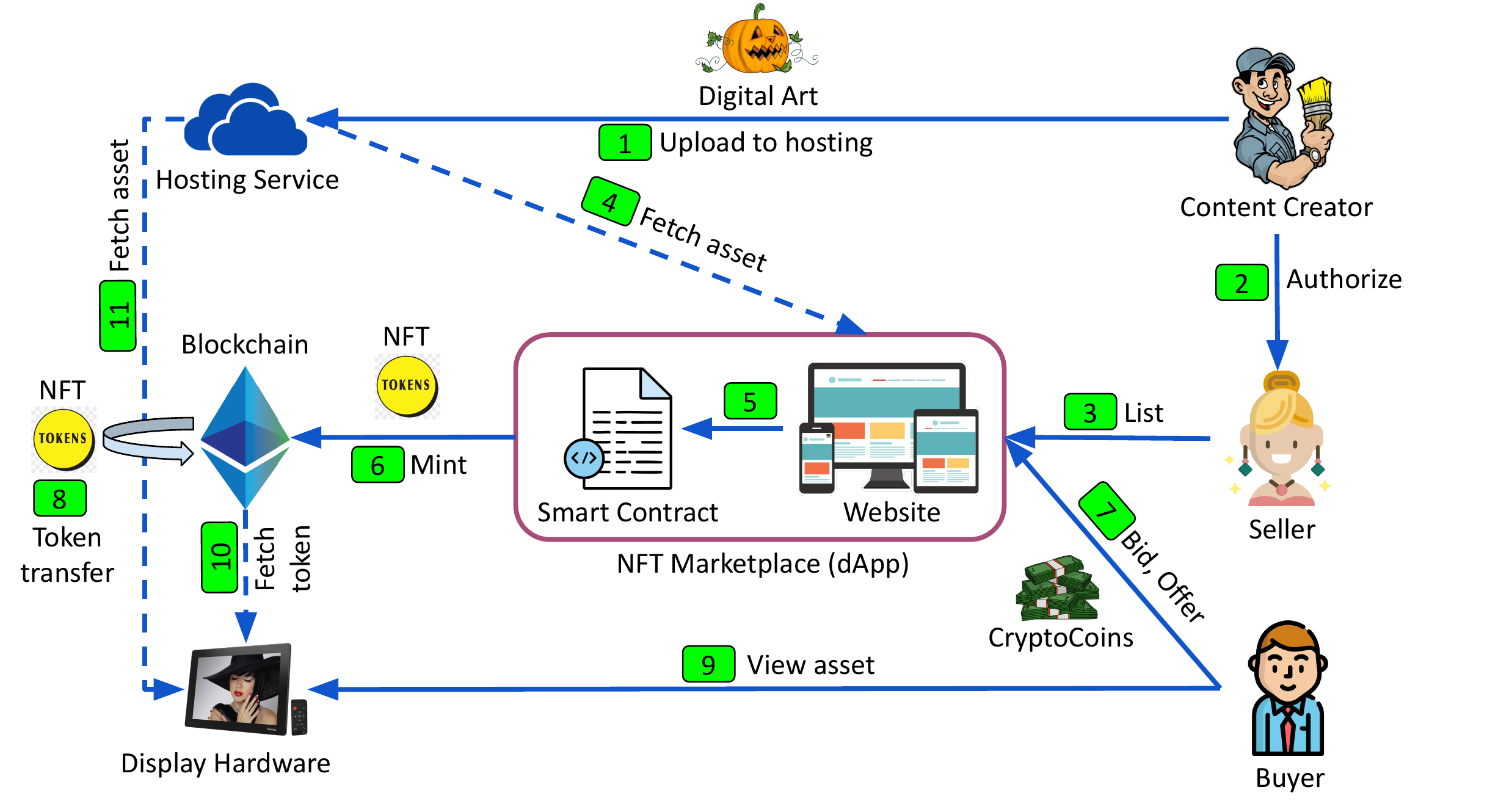}
	\vspace{-9mm}
	\caption{\small Anatomy of the NFT ecosystem showing all the marketplace actors, and their mutual interaction.
		The dotted and solid lines indicate data-centric and command-centric communication channels, respectively.}
	\label{fig:nft_ecosystem}
	\vspace{-6mm}
\end{figure}

\vspace{0.5mm}
\noindent
\textbf{InterPlanetary File System (IPFS).}
IPFS~\cite{ipfs} is a distributed, peer-to-peer, permissionless file system.
Anyone can join the IPFS overlay network.
A data item $d$ is assigned a unique immutable address, also known as content identifier (CID): $cid=H(d)$, which is the hash $H$ of the file's content $d$.
Therefore, when the content of the file changes, the CID changes as well.
The content of a file is first split into blocks.
All the storage elements, \ie, a directory, the files inside the directory, and the blocks within those files, are stored in a directed acyclic graph structure called a Merkle DAG.
IPFS maintains a distributed hash table (DHT) split across all the nodes in the network to store provider records, which locate those peers that store the requested content.
To retrieve a data item $d$, a node first looks up the providers $P(d)$ in the DHT, and then requests $d$ from the members of $P(d)$.

	\section{Anatomy of the NFT Ecosystem} 
\label{sec:ecosystem}

In this section, we provide an overview (\Fig{fig:nft_ecosystem}) of the economy that has developed around NFTs.
Specifically, we identify the actors that participate in the ecosystem and the components they interact with.

\vspace{0.5mm}
\noindent
\textbf{Users.}
NFTs are often used to sell digital collectibles and artwork, \eg, images, audio files, and videos. The  
users in the NFT ecosystem belong to one of three categories:  \textit{content creator}, \textit{seller}, and \textit{buyer}.
First, the creators create digital content and upload it \boxlabeled{1} to hosting services (an \textit{external entity}) to make the art publicly available. 
When it comes to selling the content, some creators are not technical enough to turn their art into an NFT, and put it as a token on the \bc. 
Therefore, they authorize \boxlabeled{2} sellers to mint NFTs \boxlabeled{6} and offer it on marketplaces.
In other cases, a content creator is also taking the role of the seller.
Once listed on a marketplace \boxlabeled{3}, buyers 
can buy the artwork at a listed price, make offers, or place bids \boxlabeled{7}.
If their offer is accepted or they win an auction, the NFT is transferred \boxlabeled{8} by invoking the \texttt{transferFrom()} API (\Sec{sec:background}) from the seller to the buyer to reflect the change in ownership.

\vspace{0.5mm}
\noindent
\textbf{Marketplaces.}
NFT marketplaces (\nftm) are \dapp{} platforms where NFTs (also referred to as \textit{assets}) are traded.
There are typically two main components of an \nftm---a user-facing web frontend, and a collection of \smart{}s that interact with the \bc.
Users interact with the web app, which, in turn, sends transactions to the \smart s on their behalf \boxlabeled{5}.
Primarily, there are two types of contracts: \textbf{(i)} marketplace contracts, which implement the part of the \nftm{} protocol that interacts with the \bc, and \textbf{(ii)} token contracts, which manage NFTs.
Marketplaces typically allow users to perform the following activities: \textbf{(a)} user authentication, \textbf{(b)} token minting, \textbf{(c)} token listing, and \textbf{(d)} token trading.
The token-related activities are collectively called \textit{events}.
Depending on where these events are stored, three broad types of \nftm{} protocol design are possible: \textbf{(i)} \textit{on-chain}: all the events live on the \bc.
Since every action costs gas, this design makes the \nftm{} operationally expensive for the users.
\nftm s that follow this design include \axie, \cryptopunks, \foundation, and \superrare.
\textbf{(ii)} \textit{off-chain}: the events are recorded in a centralized, off-chain database managed by the \nftm.
Users perform various activities by interacting with the web app, not the \bc, and, therefore, this design is gas-friendly.
\nifty is an example of an off-chain \nftm.
\textbf{(iii)} \textit{hybrid}: depending on their type, events are stored either on-chain or off-chain.
To ensure the integrity of the operation, on-chain and off-chain events are tied together with a cryptographic check.
\opensea and \rarible follow this model.

\vspace{.5mm}
\noindent
$\blacktriangleright$ \textbf{User authentication.}
Users first need to register with the \nftm s to access their services.
Post-registration, two different authentication workflows are possible:
\textbf{(a)} classic credentials-based (username/password), or
\textbf{(b)} signature-based.
With the latter, the user is first asked to sign a challenge string.
Then, the marketplace recovers~\cite{ecrecover} the address of the signer (user) from the elliptic-curve signature.
\opensea, \rarible, \foundation, \cryptopunks, and \superrare follow this model.
Since \ethereum private keys are essentially unguessable~\cite{eth-pvt-key}, this authentication method is generally more secure than traditional passwords (passwords are typically drawn from a limited set of characters, shorter in length, and easier to brute-force).

\vspace{0.5mm}
\noindent
$\blacktriangleright$ \textbf{Token minting.}
A token is \textit{minted} (created) \boxlabeled{6} by calling the appropriate method of the token contract, which generally complies with the \ercnonfungible{} or ERC-1155 standard.
A single token contract can manage the ownership of a number of NFTs.
Every NFT is assigned an integer called \texttt{\_tokenId}.
Therefore, an NFT is uniquely identified by the $\langle \texttt{token\_contract\_address}, \;\allowbreak \texttt{\_tokenId} \rangle$ pair on the \bc.
A ``family'' of NFTs, which are either similar, or based on a common theme, called a \textit{collection}, \eg, \cryptopunks.
An NFT can be minted in many different ways: \textbf{(a)} \textit{default contract}: the token is minted as part of a pre-deployed, designated token contract managed by the marketplace. 
\nftm s like \opensea, \foundation, \superrare, \etc, provide a default contract to hold NFTs when no custom contract is deployed by the creator.
\textbf{(b)} \textit{replica contract}: the \nftm{} itself deploys a contract on behalf of the creator to manage the collection that the NFT is a part of.
Deployed contracts have identical bytecode, but are customized through initialization parameters.
Examples of such \nftm s includes \nifty and \rarible.
Since both \textit{default} and \textit{replica} contracts are managed by the \nftm, together they are called \textit{internal} token contracts.
\textbf{(c)} \textit{external contract}: the creator independently deploys a custom contract to manage the collection, and later imports it to the marketplace.
To be interoperable with the \nftm s, external contracts must follow a well-established token standard.
Otherwise, a custom integration is needed.
\opensea and \rarible allow external contracts on their platforms.
A single token contract can manage one or more collection.
Typically, \textit{replica} or \textit{external} contracts manage a single collection, while the marketplace \textit{default} contract manages several.
In the latter case, the \nftm{} \dapp{} maintains an off-chain association between the set of \texttt{\_tokenId}s and the collection those belong to.

\vspace{0.5mm}
\noindent
$\blacktriangleright$ \textbf{Token listing.}
Once created, a seller lists their assets for sale \boxlabeled{3}.
To list an NFT on a platform, some \nftm s, \eg, \foundation, \superrare, \nifty, mandate either the seller or the entire collection (that the NFT is a part of) to be verified.
Even for the \nftm s where verification is optional, for example, \opensea, \rarible, getting an artist or a collection verified provides credibility and increases buyers' confidence.
\nftm s display special badges on verified profiles of artists and collections, which helps in building a brand, and receive preferential treatment to boost sales -- such as search priority and safe-listing to suppress safety-related alerts before the purchase.

\noindent
$\blacktriangleright$ \textbf{Token trading.}
Buyers can make offers, or place bids \boxlabeled{7} on the assets on sale.
When an offer is accepted, or an auction is settled, the \nftm{} transfers \boxlabeled{8} assets from the seller's account to the buyer's.
Usually, this is when the \nftm s charge a fee for the service they offer.
A few key aspects of the \nftm{} bidding system are discussed below: 
\textbf{(i)} \textit{Pricing protocol}: The bid price can either increase or decrease with every bid.
In an English auction, 
the bid opens at a reserve price, which is the minimum price the seller is willing to accept for an NFT.
Subsequent bids from the buyer gradually increase the price.
The NFT goes to the highest bidder.
The English auction approach is used by most \nftm s, \eg, \opensea, \foundation, and \superrare.  
In a Dutch auction,
the bid opens at a high price.
Subsequent bids from the seller gradually decrease the price.
The NFT goes to the bidder who first accepts a bid.
\axie follows the Dutch auction pattern.
\textbf{(ii)} \textit{Bid storage}:
Bids can be stored either on-chain, \eg, \cryptopunks, \foundation, \superrare, or off-chain, \eg, \nifty, \rarible, \opensea.
There are protocols, such as \wyvern used by \opensea, which keep both the sell order (listing) and the bids off-chain for gas efficiency, though the order matching and the NFT transfer happen on-chain.
Therefore, the marketplace contract cryptographically verifies the buy order against the associated sell order to prevent a malicious buyer from either buying an item that is not on sale, or tampering with an existing sell order.
\textbf{(iii)} \textit{Active bids}:
Some \nftm s disallow multiple active bids on the same asset.
For example, in \cryptopunks, \foundation, or \superrare, when a bidder outbids the current top bidder, the latter gets automatically refunded.
\textbf{(iv)} \textit{Bid withdrawal}:
Some \nftm s, such as \cryptopunks, allow the withdrawal of bids, while others, for example, \foundation, do not.
\textbf{(v)} \textit{Bid settlement}:
Bid settlement does not require seller's intervention in most cases, \ie, the asset automatically goes to the highest bidder.
However, for some \nftm s like \cryptopunks, the bid has to be explicitly accepted by the seller.

When an item is sold by a seller other than the creator, it is called a secondary sale.
Royalty is the payment made to the creator for every such secondary sale.
Before the first (primary) sale takes place, the creator specifies the royalty amount, which is then deducted from every secondary sales and given to the creator.
The deduction happens either
\textbf{(i)} on-chain, where royalty is calculated by the marketplace contract during the \textit{buy} transaction, or
\textbf{(ii)} off-chain, where the \nftm{} \dapp{} keeps track of the royalty accumulated from all the sales.

\vspace{0.5mm}
\noindent
\textbf{External entities.}
External to both \nftm s and \bc, there are services and devices that provide the necessary infrastructure for the system to work.
For example, creators store \boxlabeled{1} their artwork on web servers or storage services such as Amazon S3 or IPFS.
When buyers purchase the NFT, they can exercise their \textit{bragging right} by displaying the art on photobook-style websites or digital NFT photo-frames.
The websites, photo-frames \boxlabeled{11}, and \nftm s \boxlabeled{4} fetch tokens from the \bc{} \boxlabeled{10}, and respective artwork from those services.

	\section{Analysis approach} 
\label{sec:analysis_approach}

This paper studies scams, malpractice, and security issues in the NFT ecosystem.
In particular, we investigate the following research questions related to the three entities identified in the previous section, \ie, users, marketplaces, and external entities:
\noindent \textbf{(RQ1.)} Are there weaknesses in the way \nftm s operate today, and can those be exploited (\Sec{sec:market_place_issues})?
\noindent \textbf{(RQ2.)} How and to what extent do external entities pose a threat to the NFT ecosystem (\Sec{sec:external_entity_issues})?
\noindent \textbf{(RQ3.)} Are users involved in any fraud or malpractice resulting in the financial loss for others (\Sec{sec:user_issues})?

We used a hybrid (both qualitative and quantitative) approach to answer RQ1, and a quantitative approach for both RQ2 and RQ3.
The rest of this section discusses how we collected the data for the quantitative analysis, and we provide a rationale for choosing the specific NFT marketplaces that we examined in more detail.

\begin{table}
	\centering
	\scriptsize
	\begin{tabular}
		{R{0.22\columnwidth}	%
		L{0.10\columnwidth}		%
		L{0.12\columnwidth}		%
		L{0.12\columnwidth}		%
		C{0.05\columnwidth}		%
		C{0.05\columnwidth}}	%
		
		\toprule
		\textbf{Marketplace} & 
		\rotatebox{90}{\parbox{1cm}{\centering\textbf{Trading\\Volume}}} &
		\textbf{Assets} & \textbf{Events} &
		\rotatebox{90}{\parbox{1cm}{\centering\textbf{Asset\\Collection}}} &
		\rotatebox{90}{\parbox{1cm}{\centering\textbf{Event\\Collection}}} \\
		\midrule
		
		\rowcolor{black!10} \opensea~\cite{opensea} & \openseatradingvolume & \num{\openseaassets} & \num{\openseaevents} & A, B & A\\
		
		\axie~\cite{axie} & \axietradingvolume & \num{\axieassets} & \num{\axieevents} & B & B \\
		
		\rowcolor{black!10} \cryptopunks~\cite{cryptopunks} & \cryptopunkstradingvolume & \num{\cryptopunksassets} & \num{\cryptopunksevents} & B & B \\
		
		\rarible~\cite{rarible} & \raribletradingvolume & \num{\raribleassets} & \num{\raribleevents} & B & B, W\\
		
		\rowcolor{black!10} \superrare~\cite{superrare} & \superraretradingvolume & \num{\superrareassets} & \num{\superrareevents} & B & B\\
		
		\sorare~\cite{sorare} & \soraretradingvolume & \num{\sorareassets} & \num{\sorareevents} & B & W, B\\
		
		\rowcolor{black!10} \foundation~\cite{foundation} & \foundationtradingvolume & \num{\foundationassets} & \num{\foundationevents} & B & B\\
		
		\nifty~\cite{nifty} & \niftytradingvolume & - & - & - & - \\
		
		\bottomrule
	\end{tabular}
	\caption{\small Characteristics of the marketplace dataset.
	A: API access, W: Web scraping, B: Blockchain parsing.}
	\label{tbl:marketplaces}
	\vspace{-10mm}
\end{table}

\vspace{0.5mm}
\noindent
\textbf{Marketplace selection.}
In line with previous work~\cite{Kai21}, we use \dappradar~\cite{dappradar}, a popular tracker of dApps, to %
select the most relevant marketplaces.
We selected \marketplacesbyvolume{} out of a total of \totaldappradarmarketplaces{} marketplaces (\Tbl{tbl:marketplaces}) listed in \dappradar. This selection was based on the following two criteria:
\textbf{(a)} backed by the \ethereum blockchain, and
\textbf{(b)} the ``all-time'' trading volume is over $50$M USD as of \datacollectionstartdate.

\vspace{0.5mm}
\noindent
\textbf{Data collection.}
We collect two different types of data:
\textbf{(a)} information about the NFTs (assets), \eg, collection name, asset URI, metadata URI, \etc, traded on the different marketplaces, 
\textbf{(b)} NFT-related events, such as mint, buy, sell, auction creation, placing of a bid, acceptance of a bid, transfer, \etc, generated as a result of marketplace activity.
We provide the details of the collected data in \Tbl{tbl:data_collection}.
To conduct differential analysis for one of the studies, we needed to monitor how the  details of certain assets change over a period of time.
Therefore, we crawled the same set of assets three times with a three-month interval between two subsequent crawls: in June $2021$, September $2021$, and finally in December $2021$.
Moreover, we collected event information continuously between the June and September crawls.
We use \coingecko~\cite{coingecko} API to fetch historical prices of the cryptocoins to convert the pricing information to their equivalent USD value.

\vspace{0.5mm}
\noindent \textbf{Asset and event information}:
The first step to collect asset and event information is \textit{asset enumeration}, \ie, obtaining the list of assets traded on a marketplace.
Once enumerated, we collect the asset and event information for those assets. 
For both the steps, we employ three different strategies, subject to marketplace restrictions:

\noindent \textit{1) API access}: If a marketplace exposes an appropriate API, we use it to retrieve the list of assets and events.
Unfortunately, the APIs are often record-limited, \eg, for a specific query, \opensea{}'s API returns at most $10,000$ assets. However, the total number of assets listed on their website was \numwithsuffix{\openseatotalassets} at the time of crawling.
As a workaround, we 
generate API requests with combinations of \texttt{sort} and \texttt{filter} parameters to fetch different sets of assets with every request.

\noindent \textit{2) Web scraping}: If a marketplace does not provide an API interface, but its terms and conditions (T\&C) do not disallow scraping of their web interface, we crawl the assets and events data from the website.

\noindent \textit{3) Blockchain parsing}: If a marketplace neither provides an API nor allows web scraping, we retrieve asset and event data directly from the blockchain, if possible.
Trading activities of a decentralized marketplace are handled by \smart{}s that are well-known.
Leveraging the ABI (Application Binary Interface) of the contracts published in \etherscan, we parse historical transactions, \eg, \texttt{atomicMatch()} in case of \opensea, to retrieve asset and event details.

Our asset collection is best-effort, as it is impossible to enumerate all the listed assets in a marketplace. This is due to various reasons mentioned above, such as the absence of marketplace APIs, their rate limits, and T\&C prohibiting any crawling activity.
\Tbl{tbl:marketplaces} shows the number of assets and events collected for each marketplace, and the strategies used to collect data.
Since \nifty does not provide an API, prohibits web scraping through T\&C, and stores events off-chain, we were unable to collect data on the marketplace activities.

\vspace{0.5mm}
\noindent
\textbf{Measurement study.}
We utilize the asset and event data we collected to perform several measurement studies, which are described in the subsequent sections.
We would like to emphasize that we attain reasonable coverage, \eg, \opensea, the largest \nftm{} that accounts for $\FPeval{\v}{round(\openseaassets/(\openseaassets+\axieassets+\sorareassets+\superrareassets+\foundationassets+\cryptopunksassets+\raribleassets)*100, 2)}\v\%$ of assets in our dataset, listed \numwithsuffix{\openseatotalassets} assets in their website at the time of crawling.
We crawled \numwithsuffix{\openseaassets} assets, which is $\FPeval{\v}{round(\openseaassets/\openseatotalassets*100, 2)}\v\%$ of the size of the marketplace.
Since \opensea contributes to the most number of assets in our dataset, we use only \opensea in \Sec{sec:market_place_issues} (unless the study requires cross-\nftm{} analysis) and \Sec{sec:external_entity_issues}, as only that dataset would be representative enough to capture the extent of the issues we quantified in those sections.
However, since we measured the occurrences of trading malpractices per \nftm{} in \Sec{sec:user_issues}, we used assets from all the marketplaces.
We provide the \ethereum addresses of the major contracts relevant to our study in \Adx{sec:contract_addresses}.
	\section{Issues in NFT Marketplaces}
\label{sec:market_place_issues}

In this section, we identify weaknesses in the design of \nftm{}s, which, when abused, pose a significant risk in the form of financial loss to both the marketplaces and its users.
For this part of the study, we gathered information from public security incidents, attacks, and abuses reported on various blogs and technical reports, 
direct interactions with individual marketplaces, and marketplace documentation.
We have systematized our findings by connecting those issues with the marketplace activities discussed in \Sec{sec:ecosystem}, and then quantified, whenever possible, the prevalence/impact of those issues.
Lastly, we systematically evaluated the existence of each of the issues across all the marketplaces (\Tbl{tbl:marketplace_issues}).

\definecolor{light-gray}{gray}{0.75}
\newcolumntype{?}{!{\color{light-gray}\vrule width 0.4pt}}
\newcommand{\vthead}[1]{\makebox[1em][l]{\rotatebox{90}{#1}}}
\newcommand{\hthead}[1]{\makebox[1em][l]{\rotatebox{0}{#1}}}
\newcommand{\checkm}{\ding{52}}
\newcommand{\pie}[1]{
	\hspace{-.8ex}\hskip-\tabcolsep
	\begin{tikzpicture}[scale=0.8, baseline=-.5ex]
	\draw (0,0) circle (1ex);
	\fill[white] (0,0) circle (1ex);
	\ifthenelse{\equal{#1}{0}}{
	}{
		\fill (0,1ex) arc (90:int(#1*90+90):1ex) -- (0,0) -- cycle;
	}
	\end{tikzpicture}
	\hspace{-1.6ex}\hskip-\tabcolsep
}
\newcommand{\piel}[1]{\hspace{3pt}\pie{#1}\hspace{5pt}}
\newcommand{\pievl}[1]{\hspace{.12em}\piev{4}\hspace{5pt}}
\newcommand{\celll}{
	\hspace{-1.3ex}
	\begin{tikzpicture}[scale=0.8, baseline=.4ex]
	\fill[red!35] (0,0) rectangle (2.5ex,2.5ex);
	\end{tikzpicture}
	\hspace{-0.6ex}
}
\newcommand{\na}{{\scriptsize{n/a}}}
\newcommand{\fp}{{\cellcolor{red!35}}}

\begin{table}[!t]
\small
\centering
\setlength{\tabcolsep}{0.4em}
	\begin{tabular}{| l | c ? c ? c ? c ? c ? c ? c ? c |}
	\hline
	\multicolumn{1}{|c|}{\thead{Issues}}
	& \multicolumn{8}{c|}{\thead{Marketplaces}}
	\\
		& \rotatebox{90}{\opensea}
		& \rotatebox{90}{\axie}
		& \rotatebox{90}{\cryptopunks}
		& \rotatebox{90}{\rarible}
		& \rotatebox{90}{\superrare}
		& \rotatebox{90}{\sorare}
		& \rotatebox{90}{\foundation}
		& \rotatebox{90}{\nifty}
		\\
		\hline

		\rowcolor{black!10}\multirow{-1}{*}{\rotatebox{0}{\textbf{User authentication}}}& & & & & & & & \\
			 \rowcolor{black!10} \hspace{2mm} \textbf{U1.} Identity verification & \cross & \cross & \cross & \cross & \cross & \cross & \cross & \cross \\
			 \rowcolor{black!10} \hspace{2mm} \textbf{U2.} Two-factor authentication & \color{bulgarianrose} N & \cross & \color{bulgarianrose} N & \color{bulgarianrose} N & \color{bulgarianrose} N & \color{blue} O & \color{bulgarianrose} N & \tick \\
		\hline
		
		\rowcolor{black!5} \multirow{-1}{*}{\rotatebox{0}{\textbf{Token minting}}}& & & & & & & & \\
			\rowcolor{black!5} \hspace{2mm} \textbf{M1.} Verifiability of token contracts & \cross & \color{bulgarianrose} N & \color{bulgarianrose} N & \cross & \color{bulgarianrose} N & \color{bulgarianrose}N & \color{bulgarianrose} N & \cross \\
			\rowcolor{black!5} \hspace{2mm} \textbf{M2.} Tampering token metadata & & & & & & & & \\
			\rowcolor{black!5}\hspace{7mm} \textbf{M2.1} Changing metadata url & \color{blue} P & \tick & \cross & \cross & \color{blue} P & \color{blue} P & \cross & \cross \\
			\rowcolor{black!5}\hspace{7mm} \textbf{M2.2} Decentralized metadata & \color{blue} O & \cross & \cross & \color{blue} O & \color{blue}O & \cross & \color{blue} M & \color{blue} O \\
		\hline
		
		\rowcolor{black!10} \multirow{-1}{*}{\rotatebox{0}{\textbf{Token listing}}}& & & & & & & & \\
			\rowcolor{black!10} \hspace{2mm} \textbf{L1.} Principle of least privilege & \tick & \tick & \tick & \tick & \color{blue} P & \tick & \cross & \cross \\
			\rowcolor{black!10} \hspace{2mm} \textbf{L2.} Invalid caching & \tick &  \color{bulgarianrose} N & \color{bulgarianrose} N & \tick &  \color{bulgarianrose} N & \color{bulgarianrose} N & \color{bulgarianrose} N & \color{bulgarianrose} N \\
			\rowcolor{black!10}\hspace{2mm} \textbf{L3.} Seller / collection verification & \color{blue} O & \color{bulgarianrose} N & \color{bulgarianrose} N & \color{blue} O & \color{blue} M & \color{bulgarianrose} N & \color{blue} M & \color{blue} M \\
		\hline

		\rowcolor{black!5} \multirow{-1}{*}{\rotatebox{0}{\textbf{Token trading}}}& & & & & & & & \\
			\rowcolor{black!5} \hspace{2mm} \textbf{T1.} Lack of transparency & \cross & \cross & \cross & \cross & \cross & \cross & \cross & \tick \\
			\rowcolor{black!5} \hspace{2mm} \textbf{T2.} Fairness in bidding & \cross & \tick & \tick & \cross & \tick & \cross & \tick & \cross \\
			\rowcolor{black!5} \hspace{2mm} \textbf{T3.} Royalty and fee evasion & & & & & & & & \\
			\rowcolor{black!5}\hspace{7mm} \textbf{T3.1} Cross-platform & \cross & \cross & \color{bulgarianrose} N & \cross & \cross & \cross & \color{blue}P & \cross \\
			\rowcolor{black!5}\hspace{7mm} \textbf{T3.2} Post-sales modification & \tick & \cross & \color{bulgarianrose} N & \tick & \cross & \cross & \cross & \cross \\
		\hline
	\end{tabular}

	\vspace{1mm}
	\caption{\small Issues in the NFT marketplaces.
	{\color{blue} O}: Optional, {\color{blue}M} : Mandatory, {\color{blue} P}: Partial, {\color{bulgarianrose}N}: Not applicable, \tick: Exists, \cross: Does not exist.}
	\label{tbl:marketplace_issues}
	\vspace{-9mm}
\end{table}

\subsection{User Authentication}
\noindent
\textbf{(U1) Identity verification.}
Art in the physical world has been used in money laundering schemes~\cite{senate-report-on-art-laundering}.
NFTs might make this process easier, as trades are executed by anonymous users, and there are no physical artworks to be transported.
Identity verification is the first step to deter such criminals.
Major crypto exchanges, such as Coinbase and Binance US, are highly regulated.
To create an account with these exchanges, one needs to provide personally identifiable information (PII), \eg, name, residential address, social security number (SSN), along with supporting documents confirming these details.
Without getting the identity verified, it is either impossible to use the platform, or it can only be used with tight financial restrictions in place.
To investigate if the \nftm s impose similar regulatory restrictions, we interacted with them by creating accounts.
We discovered that no \nftm{} has made any steps towards enforcing KYC (Know Your Customer) rules nor implemented AML/CFT (Anti-Money Laundering/Combating the Financing of Terrorism) measures.
As a result, apart from being able to hide the identity, a user can create several accounts on the platform that are hard to be traced back to one single entity.

\vspace{0.5mm}
\noindent
\textbf{(U2) Two-factor authentication.}
Enabling 2FA (Two-Factor Authentication) greatly enhances the security of a password-based authentication workflow.
While traditional financial institutions like banks, brokerages, and cryptocurrency exchanges, such as \coinbase and \binance, provide 2FA as an option, it is not yet a ubiquitous option for \nftm s.
\sorare manages a user's wallet on her behalf.
As a result, an attacker who is able to login into an account can download the user's \ethereum private key associated with the wallet, and transact on behalf of her.
Though \sorare does support 2FA, it is not enabled by default.
2FA was also optional for \nifty users until the infamous hack~\cite{nifty-hack} that compromised a number of accounts in March 2021.
According to their initial assessment, none of the impacted accounts used 2FA when the hack took place.

\subsection{Token Minting}
\noindent
\textbf{(M1) Verifiability of token contracts.}
A token contract is considered ``verifiable'' if its source code is submitted to \etherscan.
Given the functional complexity of these token contracts, source code is much easier to audit than bytecode.
Verifiability of external token contracts is crucial as they can be malicious or buggy.
As an example, \opensea users complained about a malicious token contract that did not transfer tokens after purchase.
Also, to make a particular NFT valuable, sometimes NFT projects promise to circulate only a certain number (rarity) of that token.
A malicious token contract can be abused to mint more tokens than the \textit{rarity} threshold, thus dropping the token's price, which hurts the buyers.
A malfunctioning contract can burn gas without even doing any real work,
\eg, almost all \texttt{Purchase} events of the \texttt{CelebrityBreeder} contract 
failed with errors. 
Ideally, an NFT project should make the source of the underlying token contract available for public scrutiny before the NFTs are minted to make sure that they are neither malicious nor buggy.
Unfortunately, none of the \nftm s that support external token contracts mandates such contracts to be open-source.

\noindent
$\blacktriangleright$ \textbf{Quantitative analysis.}
To enumerate how abundant closed-source NFT tokens are, we queried \etherscan API for every token contract in our dataset to check if its source is present.
Out of $\num{\totalassetcontracts}$ token contracts, $\num{\totalassetcontractswithsource}$ ($\FPeval{\v}{round(\totalassetcontractswithsource/\totalassetcontracts*100,2)}\v\%$) were open-source, while the remaining $\num{\totalassetcontractswithoutsource}$ ($\FPeval{\v}{round(\totalassetcontractswithoutsource/\totalassetcontracts*100,2)}\v\%$) were closed-source, of which \num{\openseaassetcontractswithsource}  ($\FPeval{\v}{round(\openseaassetcontractswithsource/\totalassetcontractswithsource*100,2)}\v\%$) and \num{\openseaassetcontractswithoutsource} ($\FPeval{\v}{round(\openseaassetcontractswithoutsource/\totalassetcontractswithoutsource*100,2)}\v\%$) tokens belong to \opensea, respectively.

Further, we intended to evaluate if closed-source tokens are more likely to exhibit malicious behavior than open-source ones.
Since \nftm s take down NFTs when they observe or receive a report of either an abuse or a violation of the T\&C, we consider ``take-down'' as an indirect (yet strong) indication of a token being found malicious.
According to our observation, \num{\openseaassetcontractswithoutsourcetakendown} ($\FPeval{\v}{round(\openseaassetcontractswithoutsourcetakendown/\openseaassetcontractswithoutsource*100,2)}\v\%$) closed-source tokens were taken down by \opensea between June and December, which account for \$\numwithsuffix{\openseaassetcontractswithoutsourcetakendowntradingvolume} USD in trading volume.
On the contrary, only \num{\openseaassetcontractswithsourcetakendown} ($\FPeval{\v}{round(\openseaassetcontractswithsourcetakendown/\openseaassetcontractswithsource*100,2)}\v\%$) open-source tokens were taken down during the same span.

\vspace{0.5mm}
\noindent
\textbf{(M2) Tampering with token metadata.}
The metadata of a token holds the pointer to the corresponding asset.
Hence, if the metadata changes, the token loses its significance.
The \ercnonfungible{} standard for NFTs actually allows for the possibility to change a token's metadata.
However, when an NFT represents a particular asset (such as a piece of art) that is sold, changing the metadata violates the expectation of the buyer.
The location and the content of the metadata are decided at the time of minting.
A malicious creator/owner $\mathcal{A}$ can alter the metadata by manipulating either of the two post-minting: \textbf{(i)} by changing the \texttt{metadata\_url}, and \textbf{(ii)} by modifying the metadata itself.
Even if \textbf{(i)} can be disallowed at the contract level, metadata hosted on third-party (web) domains can be freely modified by $\mathcal{A}$, if she controls the domain.
This second attack can be prevented if the metadata is hosted in IPFS.
Since the URL of an object stored in IPFS includes the hash of its content, the metadata cannot be modified while retaining the same URL recorded in the NFT.

For internal token contracts, \cryptopunks, \foundation, \rarible, and \nifty offer no way to update the \texttt{metadata\_url} of an NFT.
\axie allows the creator to modify the URL at any time.
\opensea, \superrare, and \sorare allow modification by the creator until the first sale.
Since only \foundation mandates storing the metadata on IPFS, 
other \nftm{}s are susceptible to the second attack for the internal contracts.
Since no \nftm{} supporting %
external token contracts employs any check to prevent metadata tampering, both attacks are feasible.

\noindent
$\blacktriangleright$ \textbf{Quantitative analysis.}
We performed a differential analysis to determine the change in the \texttt{metadata\_url}s of external assets over a period of time.
Specifically, we monitored the \texttt{metadata\_url}s of all \num{\openseaexternalassets} external \opensea assets three times over a span of six months in an uniform interval---in June $2021$, September $2021$, and December $2021$, respectively.
Since \ercnonfungible{} metadata extensions are optional (explained in \Sec{sec:external_entity_issues}), \texttt{metadata\_url}s were completely missing for some of the assets.
Also, \opensea took down some assets during this time period, which is why 
their \texttt{metadata\_url}s could not be retrieved in the subsequent crawl.
After excluding these two kinds of assets, we were left with \num{\openseaassetswithmetadataurlinallthreecrawls} assets that had \texttt{metadata\_url}s in all three crawls.
According to our observation, the \texttt{metadata\_url}s of  \num{\openseametadataurlchangedbetweenfirstandsecondcrawl} ($\FPeval{\v}{round(\openseametadataurlchangedbetweenfirstandsecondcrawl/\openseaassetswithmetadataurlinallthreecrawls*100, 2)}\v\%$) and \num{\openseametadataurlchangedbetweensecondandthirdcrawl} ($\FPeval{\v}{round(\openseametadataurlchangedbetweensecondandthirdcrawl/\openseaassetswithmetadataurlinallthreecrawls*100, 2)}\v\%$) assets changed between the first two and the last two crawls, respectively.

\subsection{Token Listing}
\noindent
\textbf{(L1) Principle of least privilege.}
While listing an NFT, the \nftm{} takes control of the token so that when a sale is executed, it can transfer the ownership of the NFT from the seller to the buyer. 
To this end, the \nftm{} needs to be either
\textbf{(i)} \textit{the owner} of the NFT: that is, the current owner transfers the asset to an escrow account $\mathcal{E}$ during listing, or
\textbf{(ii)} \textit{a controller}: an \ethereum account $\mathcal{C}$ that can manage that specific NFT on behalf of the owner, or
\textbf{(iii)} \textit{an operator}: an \ethereum account $\mathcal{O}$ that can manage all the NFTs in that collection.
The escrow model in case \textbf{(i)} is risky because one single escrow contract/wallet $\mathcal{E}$ managed by the \nftm{} holds all assets being traded on the platform.
Therefore, the security of all assets in a marketplace depends on the security of the escrow contract or the external account that manages such contract.
This design essentially violates the principle of least privilege.
As a result, either a vulnerability in the contract or a leak of the private key of the external account could compromise the security of all the stored NFTs. 
\nifty, \foundation, \superrare follow this approach.
A safer alternative would be to adopt \textbf{(ii)} or \textbf{(iii)}, where a proxy contract $\mathcal{C}$ or $\mathcal{O}$ deployed by the \nftm{} becomes the controller of the NFT, or the operator of the entire NFT collection, respectively.
As enforced by the marketplace contract, the \nftm{} is able to transfer an NFT only when it has been put on sale and the required amount is first paid to the seller. This ensures the safety of the NFT token even in case of a marketplace hack.
If the private key of a seller (owner of an NFT) gets leaked, it can, at most, compromise the safety of that specific NFT or collection, as opposed to all the NFTs as in the case of the escrow model.

\noindent
$\blacktriangleright$ \textbf{Quantitative analysis.}
Among the \nftm s in our dataset, \foundation holds tokens in an escrow contract, while \nifty uses an Externally Owned Account (EOA) as escrow wallet.
\superrare escrows tokens only when an auction is ongoing.
The larger the number of NFTs held in escrow, the greater is the risk.
On \escrowcountingenddate, \superrare, \foundation and \nifty held $\superrareescrowednfts$, $\foundationescrowednfts$, and $\niftyescrowednfts$ NFTs in their escrow accounts, respectively.
In \Adx{sec:charts}, we show how the number of escrowed NFTs increased over time for both \nftm s.

\vspace{0.5mm}
\noindent
\textbf{(L2) Invalid caching.}
While displaying an NFT on sale, \opensea and \rarible leverage a local caching layer 
to avoid repeated requests to fetch the associated images.
If the image is updated, or disappears, the cache goes out of sync.
This could trick a buyer into purchasing an NFT for which the asset is either non-existent or different from what the \nftm{} displays using its stale cache.

\noindent
$\blacktriangleright$ \textbf{Quantitative analysis.}
To understand the potential impact of this caching issue, we measured how many \texttt{image\_url}s in our \opensea dataset are inaccessible (non-$200$ \texttt{HTTP} response code), but \opensea still serves the corresponding cached versions.
Out of total $\num{\openseaassets}$ NFTs, \texttt{image\_url}s of $\FPeval{\v}{round(\nonexistentipfsimageurl+\nonexistentnonipfsimageurl,0)}\num{\v} \FPeval{\v}{round(\v/\openseaassets*100,2)}\;(\v\%)$ tokens were inaccessible.
However, \opensea still cached  $\num{\nonexistentimageurlcachedbyopensea}\FPeval{\v}{round(\nonexistentimageurlcachedbyopensea/(\nonexistentipfsimageurl+\nonexistentnonipfsimageurl)*100,2)}\;(\v\%)$ of those inaccessible images, thus creating the illusion that the asset linked to the NFT is still alive.
One such broken collection is \godsunchained, a verified collection 
with an overall trading volume of $19.8$K \ether s.

\begin{table}[!t]
	\scriptsize
	\begin{tabular}{|lll|r|r|r|r|}
		\hline
		\multicolumn{3}{|l|}{} & \multicolumn{1}{l|}{\thead{Count}} & \multicolumn{1}{l|}{\thead{Total\\Sales}} & \multicolumn{1}{l|}{\thead{Average\\Sales}} & \multicolumn{1}{l|}{\thead{Taken\\Down}} \\ \hline
		
		\multicolumn{1}{|l|}{} & \multicolumn{1}{l|}{} & Verified & \cellcolor[HTML]{EFEFEF} \num{\openseaverifiedsellers} & \cellcolor[HTML]{EFEFEF} \$\numwithsuffix{\openseaverifiedsellerstotalsales} & \cellcolor[HTML]{EFEFEF} $\FPeval{\v}{round(\openseaverifiedsellerstotalsales/\openseaverifiedsellers, 0)}\$\num{\v}$ & \cellcolor[HTML]{EFEFEF} \openseaverifiedsellerstakendown \\ \cline{3-3}
		
		\multicolumn{1}{|l|}{} & \multicolumn{1}{l|}{\multirow{-2}{*}{Seller}} & Non-verified & \num{\openseanonverifiedsellers} & \$\numwithsuffix{\openseanonverifiedsellerstotalsales} & $\FPeval{\v}{round(\openseanonverifiedsellerstotalsales/\openseanonverifiedsellers, 0)}\$\num{\v}$ & \openseanonverifiedsellerstakendown \\ \cline{2-3}
		
		\multicolumn{1}{|l|}{} & \multicolumn{1}{l|}{} & Verified &  \cellcolor[HTML]{EFEFEF} \num{\openseaverifiedcollections} &  \cellcolor[HTML]{EFEFEF}  \$\numwithsuffix{\openseaverifiedcollectionstotalsales} & \cellcolor[HTML]{EFEFEF} $\FPeval{\v}{round(\openseaverifiedcollectionstotalsales/\openseaverifiedcollections, 0)}\$\num{\v}$ & \cellcolor[HTML]{EFEFEF} \openseaverifiedcollectionstakendown \\ \cline{3-3}
		
		\multicolumn{1}{|l|}{\multirow{-4}{*}{\rotatebox{90}{\opensea}}} & \multicolumn{1}{l|}{\multirow{-2}{*}{Collection}} & Non-verified & \num{\openseanonverifiedcollections} & \$\numwithsuffix{\openseanonverifiedcollectionstotalsales} & $\FPeval{\v}{round(\openseanonverifiedcollectionstotalsales/\openseanonverifiedcollections, 0)}\$\num{\v}$ & \openseanonverifiedcollectionstakendown \\ \cline{1-3}

	\hline
	\end{tabular}

	\vspace{1mm}
	\caption{\small Number of verified and non-verified sellers and collections, along with corresponding sales volumes.}
	\label{tbl:verified_seller_collection}
	\vspace{-10mm}
\end{table}

\vspace{0.5mm}
\noindent
\textbf{(L3) Seller and collection verification.}
Listings by verified sellers/collections are not only given preferential treatment by the \nftm{}s, but they also attract greater attention from the buyer community.
However, the verification mechanism is typically ad-hoc, 
and the final decision is at the discretion of the \nftm s.
Common requirements include sharing the social media handles of the sellers and proving their ownership, sharing contact information, collections needing to reach certain trading volume, submitting the draft files of the digital artworks, \etc{}
Marketplaces such as \foundation adopt a stricter policy by mandating verification of all the sellers on their platform.
However, there are \nftm s, \eg, \opensea, \rarible, where verification is optional.
Buyers are expected to exercise self-judgment when trading on these platforms, which, unfortunately, puts them at greater risk.

Since verification comes with financial benefits, it has been abused in different ways: \textbf{(i)} \textbf{Forging verification badge.}
Scammers forged profile pictures with an image of the verification badge overlaid on them, making the profiles appear visually indistinguishable from the verified ones at a cursory glance.
\textbf{(ii)} \textbf{Impersonation.}
Abusing weak verification procedures, scammers got their fake profiles verified by just submitting social media handles, without actually
proving the ownership of the corresponding accounts~\cite{artist-impersonation}.
\textbf{(iii)} \textbf{Wash trading.}
One of the requirements of \opensea to verify a collection is to have at least $100$ ETH in trading volume~\cite{opensea-blue-checkmark}, which is possibly hard to attain for a newly launched collection.
Historically, this requirement has incentivized people to perform \textit{wash trading}, \ie, performing fictitious trades between multiple accounts that are all under the control of the attacker, to artificially inflate sales volumes.

\noindent
$\blacktriangleright$ \textbf{Quantitative analysis.}
To highlight the economic incentive behind verification abuse, we present the number of sales and the sales volume generated by the verified and non-verified sellers and collections in \opensea in \Tbl{tbl:verified_seller_collection}.
Though only $\FPeval{\v}{round(\openseaverifiedsellers/(\openseaverifiedsellers+\openseanonverifiedsellers)*100, 2)}\v\%$ sellers and $\FPeval{\v}{round(\openseaverifiedcollections/(\openseaverifiedcollections+\openseanonverifiedcollections)*100, 2)}\v\%$ collections of \opensea are verified, the average sales per verified seller and collection are $\FPeval{\v}{round((\openseaverifiedsellerstotalsales/\openseaverifiedsellers)/(\openseanonverifiedsellerstotalsales/\openseanonverifiedsellers), 0)}\num{\v}$ and $\FPeval{\v}{round((\openseaverifiedcollectionstotalsales/\openseaverifiedcollections)/(\openseanonverifiedcollectionstotalsales/\openseanonverifiedcollections), 0)}\num{\v}$ times more that their non-verified counterparts, respectively.

Next, we measure how effective the \nftm{} verification mechanisms are in preventing abuse.
Had the verification mechanism been foolproof, then a verified collection could not be malicious, and in turn, it should never have been taken down.
However, we observed that $\FPeval{\v}{round(\openseaverifiedcollectionstakendown/\openseaverifiedcollections*100, 2)}\v\%$ of the verified and  $\FPeval{\v}{round(\openseanonverifiedcollectionstakendown/\openseanonverifiedcollections*100, 2)}\v\%$ of the non-verified \opensea collections were taken down in six months (between June and December $2021$).
This indicates that though verification attempts to reduce abuse, it fails to eliminate it completely.
The fact that the verified collections are still taken down shows that bad actors do ``slip through'' the system and verify their collections.

\subsection{Token Trading}
\noindent
\textbf{(T1) Lack of transparency.}
NFTs are asset-ownership records that should be stored on the \bc{} to allow for public verifiability.
In a decentralized setting, an NFT sale is handled by a marketplace contract $\mathcal{C}_m$ that invokes the \texttt{transfer()} API of the token contract $\mathcal{C}_t$ to transfer the token from the seller to the buyer.
Every \textit{sale} transaction and the associated \textit{transfer}, for example, the \texttt{atomicMatch()} call in case of \opensea, is visible on the \bc.
Among other things, each transaction includes the following information:
\textbf{(i)} address of the seller (current owner),
\textbf{(ii)} address of the buyer (new owner),
\textbf{(iii)} how much the NFT was sold for,
\textbf{(iv)} time of ownership transfer.
Querying for ownership has further been made easier by \ercnonfungible{} \texttt{ownerOf()} API that returns the current owner of a token.
The sales records, in conjunction with the API, permit one to reconstruct the precise sales and ownership history of an NFT.

On the other hand, if sales records and transactions are stored off-chain, it becomes impossible to verify any trades and the ownership history of an NFT. Moreover, a malicious \nftm{} can abuse this fact to forge spurious sales records to inflate the trading activity and volume.
Off-chain records are susceptible to tampering, censorship, and prone to disappear if the \nftm{} database goes down. 
Among the \nftm s we surveyed, only \nifty maintains off-chain records.
When an item is listed, \nifty takes control of the NFT by first having it transferred ($T_1$) to an escrow wallet. 
Thereafter, multiple trades can take place while \nifty holds the custody of the asset, but no sales record is ever emitted on the \bc.
If and when the owner decides to take the NFT out of \nifty, the marketplace transfers ($T_2$) the token back to the owner's account.
Since only $T_1$ and $T_2$ are visible from the \bc, no intermediate ownership and sales activity can be verified.

\vspace{0.5mm}
\noindent
\textbf{(T2) Fairness in bidding.}
\nftm s implement bidding either \textbf{(i)} on-chain, through a \smart{} that requires the bid amounts to be deposited while placing the bid, or 
\textbf{(ii)} off-chain, through the \nftm{} \dapp{} which maintains an orderbook without requiring any upfront payment.
Off-chain bidding is \textit{unfair} as it can be abused by both the \nftm{} and the users.
Since bids are not visible from the \bc, \nftm s can inflate the bid volume to create hype.
Also, placing bids is inexpensive, as there is no money transfer involved.
Therefore, such \nftm s are more susceptible to \textit{bid pollution}, a form of abuse where a large number of \textit{casual} bids are placed on items.
Since no money is locked, most of these bids are likely to fail due to a shortage of funds in the bidder's account at the time of execution.
Since on-chain bidding costs gas to place/cancel bids, it deters scammers from placing spurious bids, making abuses less frequent.
Moreover, on-chain bids reserve the bid amount upfront. Therefore, such bids invariably succeed during settlement.
In \opensea, we observed sellers complain that (attempted) sales of their items fail because the WETH balances of the winning bidders drop below the offered amounts.

\noindent
$\blacktriangleright$ \textbf{Quantitative analysis.}
Unless a bid fails due to lack of funds, an NFT gets transferred to the highest bidder at the end of the auction.
To measure the extent of bid pollution in the \nftm s, we enumerated the auctions where the highest bidder did not receive the item.
This is unfair to the seller, because the bid immediately below might be a lowball offer.
Our analysis uncovered \num{\openseatokennottransferredtohighestbidder} and \num{\raribletokennottransferredtohighestbidder}
such instances out of \num{\openseatotalauctions} and \num{\raribletotalauctions}
total auctions in \opensea and \rarible
, respectively.
We did not find any evidence of the same in the \foundation, \axie, and \superrare marketplaces.

\vspace{0.5mm}
\noindent
\textbf{(T3) Royalty distribution and marketplace fee evasion.}
If a royalty is set, every trade should earn a fee for the creator.
However, we identified ways in which users can potentially abuse the royalty implementations:
\textbf{(i)} \textbf{Cross-platform.} As explained in \Sec{sec:ecosystem}, royalty is enforced by either the marketplace contract or the \dapp, both of which are specific to an \nftm.
Also, \nftm s do not share royalty information with each other.
Therefore, royalty set on one platform is not visible from the other.
Leveraging this lack of coordination, a malicious seller can evade royalty by trading the NFT through a platform where royalty is not set, though it is set on another.
\textbf{(ii)} \textbf{Non-enforcement.} Neither royalty nor marketplace fees are  enforced in \ercnonfungible{} token contracts.
A malicious seller can thus avoid both payments by transferring (\ercnonfungible{} \texttt{transfer()}) the NFT to the buyer directly and settling the payment off-platform.
Both royalty and fees could be levied inside the \texttt{transfer} method of the token contract, though the additional logic makes the API more expensive.
\textbf{(iii)} \textbf{Post-sales modification.} \opensea and \rarible allow the creator to modify the royalty amount even after the primary sale.
Now, the royalty is calculated on the price listed by the seller.
In a potential abuse scenario, a creator can first lure a buyer $B$ by setting a low royalty and then increasing  it post-sales.
During secondary sales, $B$ may not notice this change at all, and may end up giving more royalty to the creator than initially advertised.

\noindent
$\blacktriangleright$ \textbf{Quantitative analysis.}
We discovered potential abuses of unconditional token transfer \textbf{(case ii)} to evade \nftm{} fees and royalty.
The question of evasion appears when a seller $S$ lists an NFT on a marketplace to gain popularity, but executes the trade off-platform, entirely bypassing the marketplace protocol.
There could be two possible cases.
Seller $S$ might trust the buyer $B$ and, therefore, transfers the NFT first. 
After that, $B$ settles the payment.
In the other case, the order is reversed.
For the assets listed in each \nftm, we counted the number of occurrences on the \bc{} where an address (seller) $S$ transferred the NFT to another address (buyer) $B$, and $B$ sent a payment to $S$ on-chain within $15$ minutes (before or after) the transfer transaction.
We found $\openseaonchainfeesettlement$, $\superrareonchainfeesettlement$, $\raribleonchainfeesettlement$, $\foundationonchainfeesettlement$, $\cryptopunksonchainfeesettlement$, $\sorareonchainfeesettlement$, and $\axieonchainfeesettlement$ such instances for assets listed in \opensea, \superrare, \rarible, \foundation, \cryptopunks, \sorare, and \axie, respectively.
Note that this estimate is conservative, because the payment could be made either off-chain or outside the time window that we considered for our analysis.

We also measured how often creators abuse  sellers by increasing the royalty after the primary sale \textbf{(case iii)}.
For each \opensea asset, we enumerated the ``sell'' events in increasing order of time, and counted the number of times the royalty was increased with respect to the previous sale.
We discovered \num{\royaltypostsalesmodification} instances of such royalty modifications across \num{\royaltypostsalesmodificationincollection} ($\FPeval{\v}{round(\royaltypostsalesmodificationincollection/\openseacollections*100, 2)}\v\%$) collections.

	\section{Issues Related to External Entities}
\label{sec:external_entity_issues}

The asset (picture, video) that an NFT points to must be accessible for this NFT to be ``meaningful.''
NFTs can point to assets in two ways.
If the NFT contract is \ercnonfungible{}-compliant and implements the metadata extension, then the token includes a \texttt{metadata\_url} on-chain,  which points to a metadata record (JSON).
This record, in turn, includes an \texttt{image\_url} field that points to the actual digital asset.
Many older tokens, on the other hand, are not standard-compliant and do not contain any on-chain \texttt{image\_url}. 
Instead, they use some ad-hoc, off-chain scheme to link to an asset.
For such NFTs, \nftm{}s implement custom support so that they can generate valid image URLs.
Since both the metadata record and the asset are  stored off-chain, those do not enjoy the same guarantee of immutability as the NFT itself.
When any URL becomes inaccessible, that breaks the link between the NFT and the corresponding asset.
In practice, the URLs frequently point to a distributed storage service, \eg, IPFS, or centralized storage, \eg, a web-domain or Amazon S3 bucket.
For IPFS URLs, if the NFT owner is aware, she can keep the NFT ``alive'' by pinning the resource (\ie, storing it persistently).
Even that could also be problematic, because NFTs do not store the hash value of the actual resource but rather store URLs that point to an IPFS gateway web service. If the gateway becomes unavailable, the NFT ``breaks.''
In general, NFTs that include URLs that point to domains outside the control of the NFT owners risk getting invalidated when the corresponding domains go away.

\noindent
$\blacktriangleright$ \textbf{Quantitative analysis.}
We performed an analysis to quantify the number of \opensea NFTs that were ``lost'' due to the reasons outlined above.
As of \datacollectionstartdate, out of our \num{\openseaassets} assets from \opensea, there were only $\FPeval{\v}{round(\nonexistentipfsmetadataurl+\existentipfsmetadataurl+\nonexistentnonipfsmetadataurl+\existentnonipfsmetadataurl, 0)}\num{\v}$ assets with a valid \texttt{metadata\_url} field.
Querying \opensea's API, we obtained
$\FPeval{\v}{round(\nonexistentipfsimageurl+\existentipfsimageurl+\nonexistentnonipfsimageurl+\existentnonipfsimageurl, 0)}\num{\v}$ assets with non-empty \texttt{image\_url} fields.
The remaining 3,860,607 assets did not have an \texttt{image\_url} field, which means that they are hosted directly on \opensea (content creators have the option to leave the image URL field empty, in which case \opensea handles the hosting).
We first check whether the image and metadata URLs point to resources hosted on IPFS.
Next, we check whether the URLs are still accessible.
To this end, we perform an \texttt{HTTP} \texttt{HEAD} query.
If the query returns with a response code other than $200$ (\texttt{OK}), we perform an \texttt{HTTP} \texttt{GET} query next.
If that also returns a non-$200$ response code, we mark that URL as \textit{inaccessible}.
We take this two-step approach to optimize for performance, and not to generate false negatives due to web servers that do not respect \texttt{HEAD} queries.
Also, the servers hosting the assets could be offline at the time of testing, but later come back up online.
To account for this possibility, we repeated the above URL-check three times in a span of $15$ days.
Only the assets marked as \textit{inaccessible} in the previous attempt were tested for accessibility each time.
An asset is finally marked as \textit{inaccessible} only if all three attempts agree.

\begin{figure}[t]
	\centering
	\includegraphics[width=0.8\linewidth]{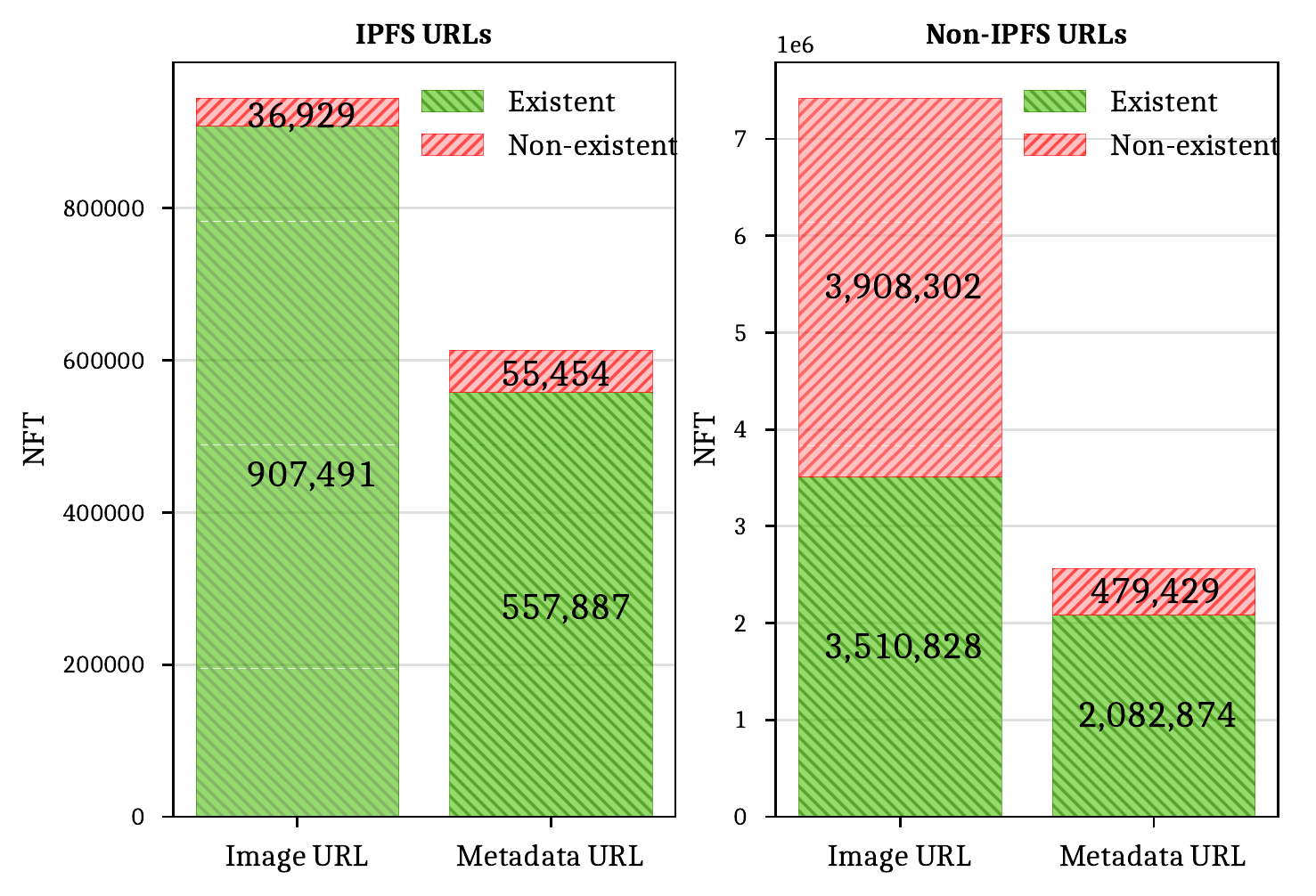}
	\vspace{-5mm}
	\caption{\small Validity of image and metadata URLs}
	\label{fig:hosting}
	\vspace{-8mm}
\end{figure}

\Fig{fig:hosting} reports our findings.
Two important observations are:
\textbf{(i)} Only $\FPeval{\v}{round(\nonexistentipfsimageurl/(\nonexistentipfsimageurl+\existentipfsimageurl)*100,2)}\v\%$ of the assets (images) and $\FPeval{\v}{round(\nonexistentipfsmetadataurl/(\nonexistentipfsmetadataurl+\existentipfsmetadataurl)*100,2)}\v\%$ of metadata records hosted on IPFS have disappeared in our dataset between June and December; as expected, NFTs hosted on IPFS are less likely to disappear than those hosted on non-IPFS domains.
\textbf{(ii)} Though IPFS is supposed to be more resilient to disappearance of the assets, a majority of asset URLs ($\FPeval{\v}{round((\nonexistentnonipfsimageurl+\existentnonipfsimageurl)/(\nonexistentipfsimageurl+\existentipfsimageurl+\nonexistentnonipfsimageurl+\existentnonipfsimageurl)*100,2)}\v\%$) as well as metadata URLs  ($\FPeval{\v}{round((\nonexistentnonipfsmetadataurl+\existentnonipfsmetadataurl)/(\nonexistentipfsmetadataurl+\existentipfsmetadataurl+\nonexistentnonipfsmetadataurl+\existentnonipfsmetadataurl)*100,2)}\v\%$) are hosted on non-IPFS domains.
Looking at all lost NFTs, they have generated a staggering amount of \$\num{\lostnfttradingvolume} USD in revenue from \num{\lostnftsalestxvolume} transactions.
Not only that, due to the caching issue we discussed in \Sec{sec:market_place_issues}, it is very well possible that a fraction of them are still in circulation.
As this analysis shows, persistence is a pressing issue in the NFT space.

	\section{Fraudulent User Behaviors}
\label{sec:user_issues}

In this section, we study the impact of various fraudulent user activities that occur in \nftm s.
In particular, we look at counterfeit NFT creation as well as trading malpractices, such as wash trading, shill bidding, and bid shielding.
In \Adx{sec:user_issues_extended}, we then cover a few more types of malicious activities that were reported in blogs and articles.

\subsection{Counterfeit NFT Creation}
The authenticity of an NFT is endorsed by the  \smart{} managing the collection.
Therefore, to ensure that the token one is buying is legitimate, buyers are advised to verify the contract address of the collection from official sources, \eg, the project's web page, before making a purchase.
Unfortunately, buyers are not always aware of the existence of counterfeits, or of how they can verify an NFT's authenticity. Instead, they only rely on the names and visual appearances of items in the marketplaces. This makes it possible for malicious users to offer ``fake'' NFTs. We observed the following types of counterfeits: 

\noindent
\textbf{(i) Similar collection names.}
There are fake NFTs that use the name of a collection or individual piece that resembles the original (victim) one.
A common trick is to substitute ASCII characters in the original name with non-ASCII characters that look alike.
To prevent such abuse, \opensea restricts users from using popular collection names and certain special characters. 
Still, it is often possible to circumvent these limitations, \eg, by adding a dot(.) at the end of the name or substituting an upper-case character with a lower-case one, \eg, a fake of ``CryptoSpells'' collection used the name ``Cryptospells.''
Moreover, restrictions can cause problems for legitimate users, \eg, French users complained about not being able to use the accented characters in collections.

\noindent
\textbf{(ii) Identical image URLs.}
Some fake NFTs point to existing assets, \ie, they simply copy the \texttt{image\_url}s of legitimate NFTs.
For example, \cryptopunks is a well-known collection.
Of course, nothing prevents a scammer from deploying her own token contract on the \bc{} and mint tokens that point to \cryptopunks.
A buyer who just looks at the appearance of items in a collection will see the \cryptopunks images and might mistake the NFTs for the originals. 

\noindent
\textbf{(iii) Similar images.}
Instead of copying the \texttt{image\_url}, a scammer might copy the digital asset and then mint an NFT that points to this copy.
As of now, no \nftm{} runs any similarity check to detect if a media file has already been used by other NFTs.

\noindent
$\blacktriangleright$ \textbf{Quantitative analysis.}
We looked for each type of counterfeits present in the \opensea dataset comprising of \num{\openseaassets} NFTs spread across \num{\openseacollections} collections.

\textbf{(i)} To check for (potential) counterfeits that abuse similar collection names, we compute the \textit{Levenshtein distance}, an \textit{edit distance} metric between pairs of collection names (strings).
Since a shorter distance indicates greater similarity, we considered a maximum distance of $2$ characters, which means that we only consider collection names as similar if they differ in at most two characters.
We considered \num{\totalcollectionsgreaterthaneightchars} collections that have names longer than $7$ characters, and a minimum of $10$ NFTs in it (collections with fewer NFTs could be insignificant) to avoid spurious matches.
Given that it is  more beneficial to imitate verified collections, we only considered collection pairs that include one verified collection (and the other one is considered to be its \textit{replica}).

Our analysis found \num{\totalsimilarverifiedcollectionpairs}  collection pairs with similar names.  
We noticed that the names of most of the replica collections were minor modifications of the names of the respective verified collections, for example, pluralizing a noun, adding whitespace at hard-to-notice positions, \etc, which indicates a potential intent to mislead.
We then randomly picked  \num{\similarcollectionnamepairsmanuallyverified} pairs and checked if those replica collections indeed contain images that are similar to the verified ones and that could mislead buyers.
Since judging the similarity visually could be subjective, two researchers independently performed the assessment, and a pair was marked ``visually similar'' only if both the decisions agreed.
We discovered \num{\similarcollectionnamepairsfoundsimilar} such collections, which we reported to \opensea requesting a take-down.
Moreover, we identified an additional  \num{\similarcollectionnamepairstakendown} collections that were already taken down by \opensea (which indicates wrongdoing) between June and December $2021$.

\textbf{(ii)} To check for  counterfeits that leverage  identical \texttt{image\_url}s, we first collected $\FPeval{\v}{round(\nonexistentipfsimageurl+\existentipfsimageurl+\nonexistentnonipfsimageurl+\existentnonipfsimageurl, 0)}\num{\v}$ \texttt{image\_url}s comprising of $\FPeval{\v}{round(\nonexistentipfsimageurl+\existentipfsimageurl, 0)}\num{\v}$ IPFS, and $\FPeval{\v}{round(\nonexistentnonipfsimageurl+\existentnonipfsimageurl, 0)}\num{\v}$ non-IPFS URLs from our dataset.
Objects on IPFS are accessed through IPFS gateways, which are web services.
An IPFS URL is typically of the form: \texttt{http(s)://<gateway>/<ipfs\_hash>}.
Any \texttt{gateway} can be used to access the object pointed to by \texttt{<ipfs\_hash>}.
Therefore, we pre-processed those URLs to extract only the hash component.
In the last step, we performed a string comparison between every pair of IPFS hashes and non-IPFS URLs, which reported \num{\nftwithidenticalurlsipfs} and \num{\nftwithidenticalurlsnonipfs} identical IPFS, and non-IPFS URLs with at least one duplicate, respectively.

\textbf{(iii)} To find potential counterfeits due to image similarity, we crawled the images pointed by the \texttt{image\_url} for all NFTs in our dataset.
Since the individual assets linked to NFTs can be very large, we decided to focus on downloading just the smaller resolution version of an asset generated and cached by \opensea. We then used the \textit{perceptual} algorithm~\cite{perceptual-hash} of \imagehash~\cite{imagehash}, a popular ($2.1$K \github stars) image hashing tool, to compute a ``fuzzy'' hash that is tolerant to small perturbations of the images.
Lastly, we compare every pair of hashes to find similar images.
We refrain from comparing hashes of the images that are part of the same collection, as they are likely similar (but not counterfeits). 
We downloaded \num{\totalnftimages} images, and we discovered \num{\nftwithsimilarimages} hash collision pairs.
We randomly picked $100$ such pairs, and manually verified that $\truepositivepctsimilarimages\%$ of those image pairs are indeed visually identical.

\begin{figure*}
	\centering
	\begin{minipage}{.3\textwidth}
		\centering
		\includegraphics[width=\linewidth]{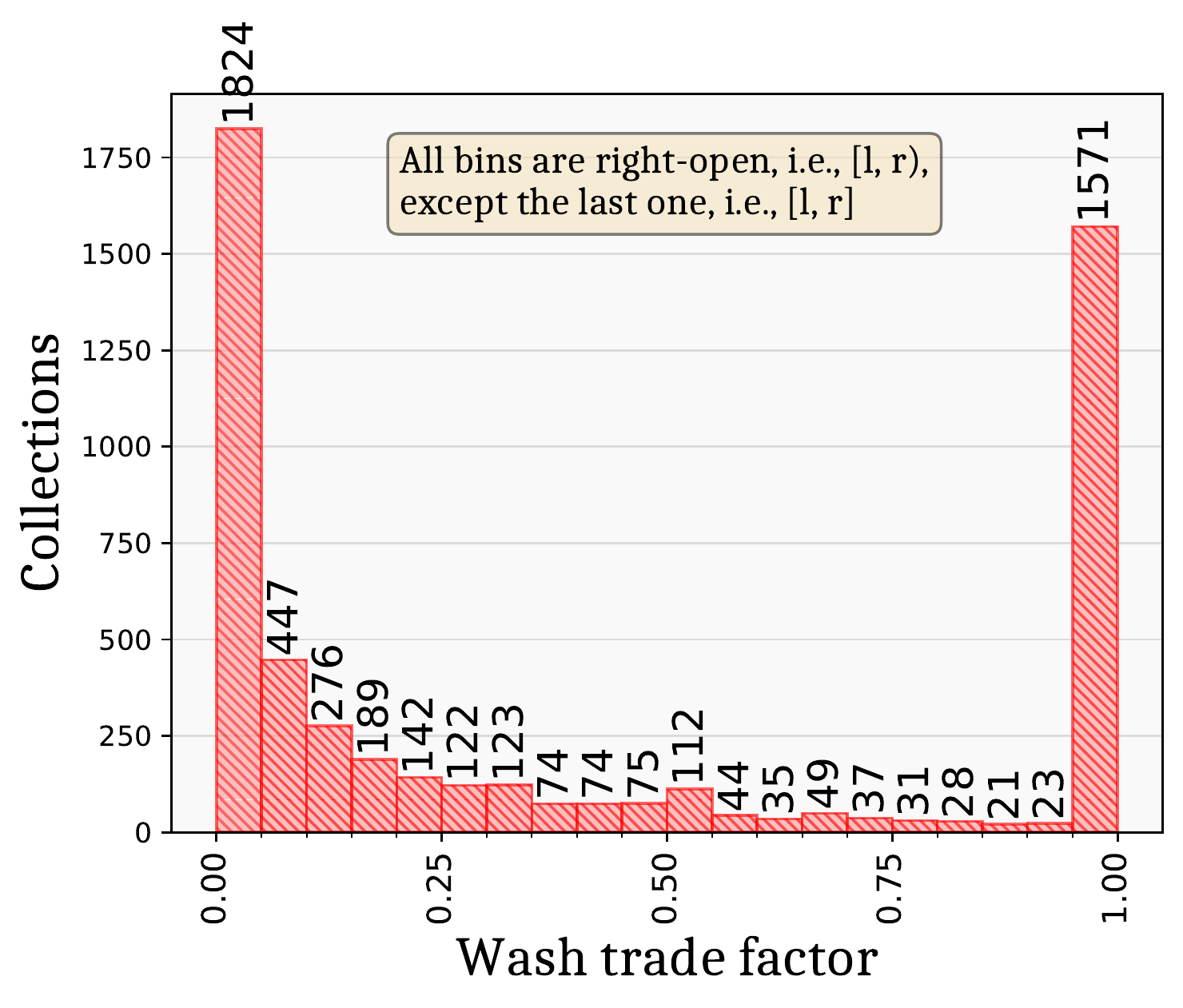}
		\caption{\small Distribution of wash trading factors across collections}
		\label{fig:wash_trade}
	\end{minipage}
	\begin{minipage}{.3\textwidth}
		\centering
		\includegraphics[width=\linewidth]{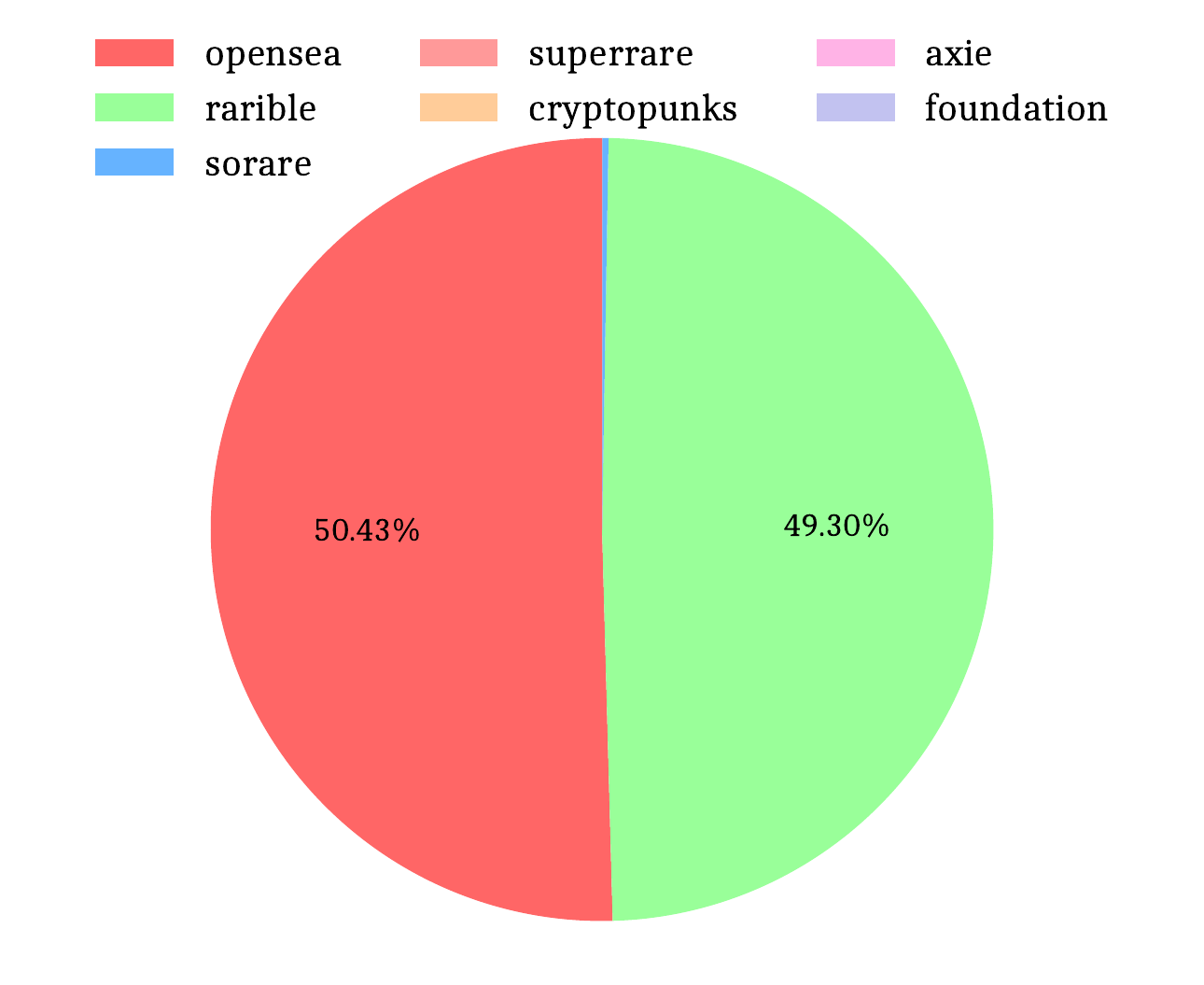}
		
		\caption{\small Relative volumes of wash \\trading in different marketplaces}
		\label{fig:wash_trade_by_marketplace}
	\end{minipage}
	\begin{minipage}{.3\textwidth}
		\centering
		\includegraphics[width=\linewidth]{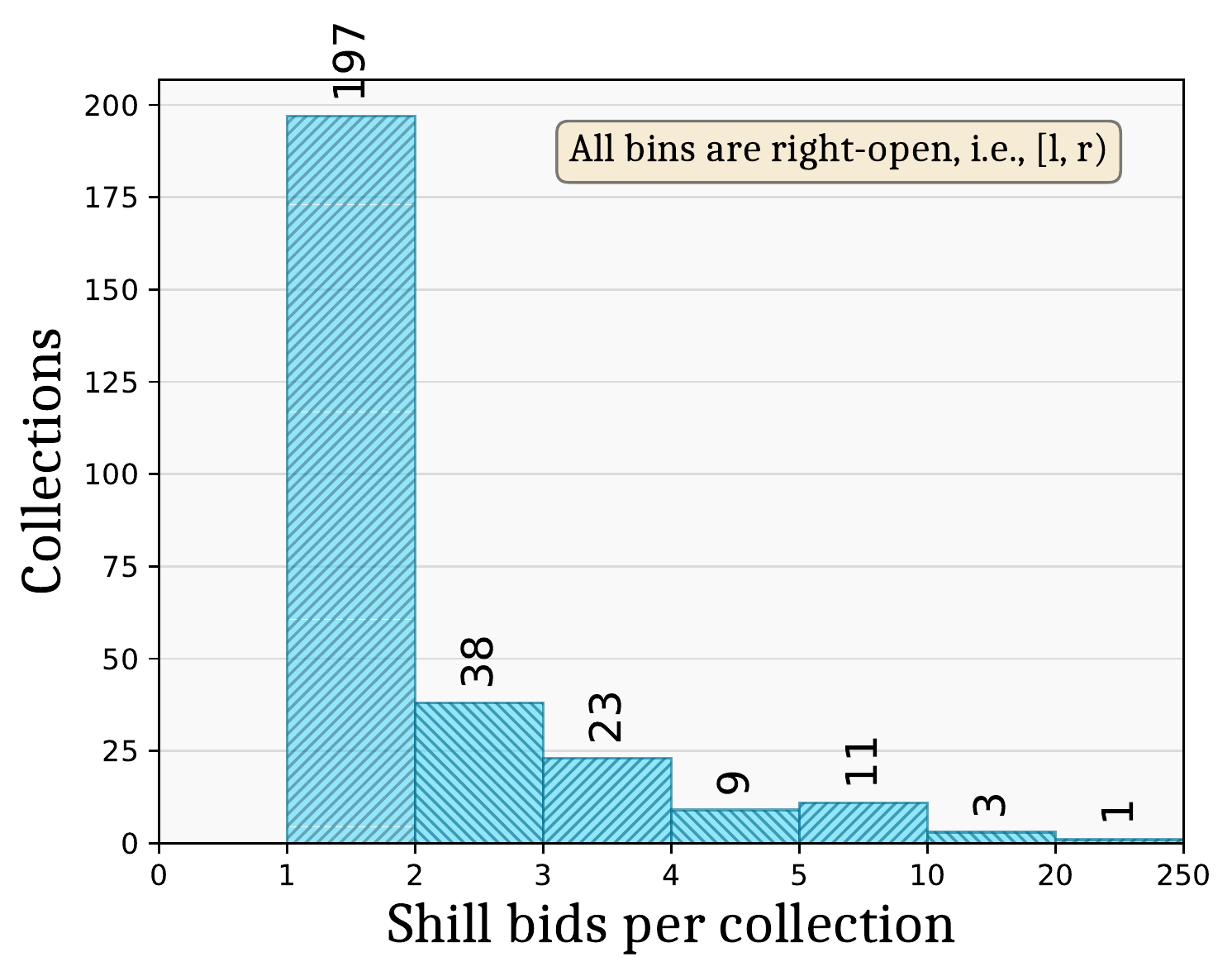}
		\caption{\small Distribution of shill \\bidding across collections}
		\label{fig:shill_bid}
	\end{minipage}%
	\vspace{-2mm}
\end{figure*}

\subsection{Trading Malpractices}
\label{sec:trading_malpractices}

In this section, we explore illicit trading practices, specifically, wash trading, shill bidding, and bid shielding~\cite{trading_malpractice, wash_trading1, shill_bidding1}.
We first discuss how these malpractices are relevant in the context of \nftm s, and then build heuristic models to detect such attacks.
Finally, we apply these models to all  $\FPeval{\v}{round(\openseaassets+\axieassets+\cryptopunksassets+\raribleassets+\superrareassets+\sorareassets+\foundationassets,0)}\num{\v}$ assets and $\FPeval{\v}{round(\openseaevents+\axieevents+\cryptopunksevents+\raribleevents+\superrareevents+\sorareevents+\foundationevents,0)}\num{\v}$ events we collected (\Sec{sec:analysis_approach}). The goal is to measure the extent and impact of these trading activities on the top \marketplacesanalyzedfortradingmalpractices{} \nftm s.

\begin{figure}[htbp]
	\vspace{-5.1mm}
	\[\begin{array}{rlll}
	\mathsf{sale}(u_1,a,p,t,u_2) & :- & \text{$u_1$ sold $a$ to $u_2$ at price $p$ }\\
	& & \text{ at time $t$ }\\
	
	\mathsf{auction}(u, p, t, id, a) & :- & \text{$u$ started auction with id $id$}\\
	&& \text{at time $t$ with starting price $p$}\\				
	
	\mathsf{bid}(u, p, t, id, a) & :- & \text{$u$ placed bid $p$ on $a$ at time $t$ }\\
	& & \text{on an auction with id  $id$}\\
	
	\mathsf{cancel\_bid}(u, p, t, id, a) & :- & \text{$u$ canceled bid $p$ on $a$ at $t$}\\
	& & \text{on an auction with id $id$}\\
	
	\mathsf{win}(u,p,t,id, a) & :- & \text{$u$ won auction $id$ on $a$}\\
	&& \text{at time $t$ with price $p$}\\
	
	\mathsf{paid}(u_1, e, u_2) & :- & \text{$u_1$ transfered $e$ ethers to $u_2$}\\
	
	\mathsf{transfer}(u_1,a,u_2) & :- & \text{$u_1$ transfered $a$ to $u_2$}\\
	\end{array}\]
	\vspace{-5mm}
	\caption{Relationships in graphs $\mathcal{G}_s, \mathcal{G}_b, \mathcal{G}_p, \mathcal{G}_t$.}
	\label{fig:relations}
	\vspace{-7mm}
\end{figure}

\vspace{0.5mm}
\noindent
\textbf{Data modeling.}
From the event data and the \ether{} flows collected from \bc{} transactions, we extract actions (such as transfers, sales, bids, ...) that operate on NFTs. \Fig{fig:relations} shows the types of predicates (actions) that we record for users $u$, assets $a$, auctions $id$ and prices $p$.
These predicates capture relationships that we use to build four different graphs: A sales graph $\mathcal{G}_s$ ($\mathsf{sale}$), a bidding graph $\mathcal{G}_b$ ($\mathsf{auction}, \mathsf{bid}, \mathsf{cancel\_bid}, \mathsf{win}$), a payment graph $\mathcal{G}_p$ $(\mathsf{paid})$, and an asset transfer graph $\mathcal{G}_t$ $(\mathsf{transfer})$.
$\mathcal{G}_b$ contains two types of nodes: users ($u$) and assets ($a$), and directed edges from $u$ to $a$ annotated with property tuples of the form $(p, t, id)$.
All of $\mathcal{G}_s$, $\mathcal{G}_p$, and $\mathcal{G}_t$ contain only one type of node: users ($u$), and directed edges from $u_1$ to $u_2$.
Edges in $\mathcal{G}_s$, $\mathcal{G}_p$, and $\mathcal{G}_t$ are annotated with property tuples of the form $(a, p, t)$, $(e)$, and $(a)$, respectively.

\subsubsection{Wash Trading}
\label{sec:wash_trading}

In wash trading, the buyer and the seller collude to artificially inflate the trading volume of an asset by engaging in spurious trading activities.
In \nftm s, users wash trade to either create the illusion of demand for a specific asset, artist, \etc, or to inflate metrics that are of their financial interest, such as getting a profile/asset verified, or collecting rewards. 
For example, \rarible users are incentivized by \$RARI governance tokens where the more a user spends, the more tokens they receive~\cite{rari_rewards_tweet}.
It is suspected that many high-value NFT sales related to popular projects such as \cryptokitties~\cite{cryptokitties} and \decentraland~\cite{decentraland} are instances of wash trading~\cite{wash_trading1}.

\vspace{0.5mm}
\noindent
\textbf{Detection.}
In the NFT space, wash traders primarily intend to increase the sales volume of NFT collections.
To detect wash trading, given a set of assets $\mathcal{A} = \{a_1, a_2, ..., a_n\}$ that are part of a collection, we check for a set of users (addresses) $\mathcal{U} = \{u_1, u_2,..., u_m\}$ who heavily trade those assets with each other.
We assume a limited number of colluding users to make the problem tractable (we use an empirical threshold of \washtradecomponentsize{} users).
In other to generate wash trades, these users repeatedly trade that set of assets among them, which often results in cycles in the sales graph $\mathcal{G}_s$.
Hence, we check $\mathcal{G}_s$ for the existence of cyclic relationships among these users.
In a strongly connected component (SCC) of a graph, there exist paths between all pairs of vertices.
Therefore, this type of wash trade can be detected~\cite{Friedhelm21} by checking if two users: $u_1$ and $u_2$ appear in any SCC of the sales graph $\mathcal{G}_s$.
In other words, if $\mathsf{SCC}(u_1, u_2, \mathcal{G}_s)$ holds, it means that both the users are involved in round-trip trades, \ie, there exist either direct, or indirect sale relations between them in both the directions.
Now, two users being a part of an SCC can be accidental, and does not indicate the frequency of trades between them.
However, in a wash trade, users are involved in frequent sales.
Therefore, we only consider SCCs where the number of sale relationships between every two intermediate users is above (indicating `heavy' trading volume) an empirically determined threshold ($\epsilon$).
We use $\epsilon=\washtradesccfrequencythreshold$ in our analysis.

However, bad actors can come up with more intricate strategies to conceal apparent connections so that such simple detection can be evaded.
We manually analyzed the \bc{} transactions history and found two evasion strategies that would throw off the prior analysis.
In the first case, when we investigated an otherwise legitimate-looking sale relation $u_i \rightarrow u_j \rightarrow u_k$, we realized that both $u_j$ and $u_k$ are funded (\ether{} transfer) by the same ``parent'' user $u_i$.
We capture this case by checking if two users: $u_1$ and $u_2$ appear in any weakly connected component (WCC) of the payment graph $\mathcal{G}_p$. 
In other words, if $\mathsf{WCC}(u_1, u_2, \mathcal{G}_p)$ holds, it means that direct or indirect \ether-flow exists between those two users in either direction.
In the second case, for a sale relation $u_i \rightarrow u_j \rightarrow u_k$, we identified multiple unconditional asset transfers (\ercnonfungible{} \texttt{transfer()}) from $u_i$ to $u_k$, giving a strong indication of a close tie between those users.
We capture this case by checking if two users: $u_1$ and $u_2$ appear in any WCC of the transfer graph $\mathcal{G}_t$.
In other words, if $\mathsf{WCC}(u_1, u_2, \mathcal{G}_t)$ holds, it means that direct or indirect unconditional asset transfer relationships exist between those two users in either direction.

To summarize, our model considers any $\mathsf{sale}(u_1,\_,\_,\_,u_2)$ relation a potential wash trade if: $\mathsf{SCC}(u_1,u_2, \mathcal{G}_s) \lor \mathsf{WCC}(u_1,u_2, \mathcal{G}_t) \lor \mathsf{WCC}(u_1,u_2, \mathcal{G}_p)$.

\vspace{0.5mm}
\noindent
$\blacktriangleright$ \textbf{Quantitative analysis.}
We detected \num{\washtradedetected} instances of wash trading that generated \$\num{\totalwashtradevolume} USD in trading volume across \num{\collectionswashtraded} collections involving \num{\userswashtraded} users in all \nftm s except \axie, \foundation, and \cryptopunks.
Moreover, out of \num{\totalcollections} collections in our dataset, only \num{\totalcollectionswithmorethantwoktxvolume} collections had more than \$$2$K in trading volume, out of which \num{\collectionswashtradedmorethantwoktxvolume} $(\FPeval{\v}{round(\collectionswashtradedmorethantwoktxvolume/\totalcollectionswithmorethantwoktxvolume*100, 2)}\v\%)$ collections show signs of wash trading.

We define \textit{wash\_trade\_factor} (WTF) as the fraction of the total trading volume of a collection generated by wash trading, \ie, if WTF is $1$, then all the trades are wash trades.
In \Fig{fig:wash_trade}, we show the distribution of the \textit{wash\_trade\_factor} across  collections where wash trading has been detected.
Of all the wash traded collections, \num{\collectionswithlessthanfivepctwashtrades} $(\FPeval{\v}{round(\collectionswithlessthanfivepctwashtrades/\collectionswashtraded*100,2)}\v\%)$ collections had less than $5\%$ ($\text{WTF} < 0.05$) of the trades generated by wash trades.
Interestingly, we discovered \num{\collectionsmorethannintyfivepctwashtrades} $(\FPeval{\v}{round(\collectionsmorethannintyfivepctwashtrades/\collectionswashtraded*100,2)}\v\%)$ collections which were heavily abused, because more than $95\%$ of all of their trades are wash trades, totaling  \$\num{\totalwashtradevolumebycollectionsmorethannintyfivepctwashtrades} USD in the trading volume.
\Fig{fig:wash_trade_by_marketplace} shows the relative volumes of wash trades that have happened in different \nftm s.
Though nearly equal volume of wash trades were discovered in both \rarible ($\FPeval{\v}{round(\washtradevolumerarible/(\washtradevolumeopensea+\washtradevolumeaxie+\washtradevolumecryptopunks+\washtradevolumefoundation+\washtradevolumerarible+\washtradevolumesorare+\washtradevolumesuperrare)*100, 2)}\v\%$) and \opensea ($\FPeval{\v}{round(\washtradevolumeopensea/(\washtradevolumeopensea+\washtradevolumeaxie+\washtradevolumecryptopunks+\washtradevolumefoundation+\washtradevolumerarible+\washtradevolumesorare+\washtradevolumesuperrare)*100, 2)}\v\%$), given that the overall trading volume of \opensea is $21$ times more (\Tbl{tbl:marketplaces}) than that of \rarible, it seems that wash trading is significantly more frequent in \rarible than \opensea.
Our finding is also corroborated by  discussions we saw on \rarible{} \discord, which indicates a heavy amount of past wash trading incidents as malicious users attempted to secure \$RARI tokens.

\vspace{0.5mm}
\noindent
$\blacktriangleright$ \textbf{Manual analysis.}
In our analysis, the size of a connected component represents the number of addresses involved in a wash trade.
We observed that $\FPeval{\v}{round(\washtradinginstanceswithlessthanequaltotencomponentsize/\washtradedetected*100, 2)}\v\%$ (\num{\washtradinginstanceswithlessthanequaltotencomponentsize}  of \num{\washtradedetected}) of the reports had a component size at most $10$.
Therefore, for our manual analysis, we randomly selected  \num{\washtradingmanuallyanalyzed} reports, and checked whether one of the following two conditions holds:
\textbf{(i)} if $a$ addresses are involved in $t$ transactions on $n$ NFTs, then both $t \geq 2a$ and $n \ll t$ need to hold.
The intuition is that if a set of users ``heavily'' trades on only a small number of assets, then those are likely to be wash trades.
Alternatively, \textbf{(ii)} the addresses involved in trading are all funded by a common, on-chain funding source.
This is true when the ``supposed'' buyers, in reality, are all funded directly by the seller, or by a seller-controlled address, before making (pseudo) purchases.
If one of these two conditions holds, we consider a detected wash trade instance as a true positive.
We determined all the sampled instances as true positives.

\vspace{0.5mm}
\noindent
\textbf{Limitation.}
\ethereum mixers (informally ``tumblers''), such as Bitmix~\cite{bitmix}, ETH Mixer~\cite{eth-mixer}, and Tornado Cash~\cite{tornado-cash}, are anonymity services that help to conceal the true source of a payment by breaking the link between the receivers and the sender of the funds.
Specifically, these services accept \ether s from a user, and either route it to a \smart, or relay it through a complex, large network of addresses by splitting the amount into a number of micro-transactions; essentially mingling that fund with hundreds of other users.
Since our wash trade detection strategy leverages information about \ether{} flows between two addresses, mixers can lead to false negatives.

\subsubsection{Shill Bidding}
\label{sec:shill_bidding}

Shill bidding is a common auction fraud where a seller artificially inflates the final price of an asset either by placing bids on her own asset, or colluding with other bidders for placing spurious bids with increasingly higher bid amounts.
This can lead to honest bidders paying higher prices than they would have otherwise.
With high-value bids on assets becoming increasingly common, it is suspected that many sales suffer from artificial price inflation~\cite{shill_bidding1}.

\vspace{0.5mm}
\noindent
\textbf{Detection.}
Detecting shill bidding is difficult when looking at a single auction in isolation. 
It becomes even harder when malicious users take turns, placing bids on each others' auctions so that the seller-bidder relation changes.
In this paper, we only consider the simple case where a specific user repeatedly places bids in auctions, yet never (or rarely) purchases anything. Moreover, we check whether there is some relationship between this user and the seller. Thus, our findings should be viewed as a lower bound on the actual number of shill bidding occurrences in \nftm s.
Our detection mechanism draws on our insight from the manual analysis of \nftm{} activities 
and prior work~\cite{Trevathan2008}.

Let $\mathsf{bid}(u_b, p_i, t_i, id, a)$ denote the $i$-th bid placed by user $u_b$ with amount $p_i$ at time $t_i$ on asset $a$ in an auction with id $id$ created by the seller $u_s$. Then, user $u_b$ is a shill bidder if:

\noindent
\textbf{Rule 1.} $u_b$ places at least $n$ bids on an asset $a$ auctioned by $u_s$ with monotonically increasing bid amounts. That is, $\forall i \in [1,n],  \mathsf{bid}(u_b, \allowbreak p_i, t_i, id,a)$, the following holds: $\forall i, \forall j, \;\; t_i > t_j \implies p_i > p_j$

\noindent
\textbf{Rule 2.} $u_b$ never buys the asset $a$, \ie, $\mathsf{win}(u_b,\_,\_,id, a)$ is $\mathsf{false}$ .

\noindent
\textbf{Rule 3.} $u_b$ has limited buying/selling activity, \ie, $|\{\mathsf{sale}(u_b,\_,\_,\_,\_)\} \allowbreak \cup \{\mathsf{sale}(\_,\_,\_,\_,u_b)\}| < \sigma$, where $\sigma$ is an empirically determined threshold.
We set $\sigma=10$ for our analysis.

\noindent
\textbf{Rule 4.} $u_b$ is ``connected'' to the seller $u_s$ either through \ether flows ($\mathcal{G}_p$) or asset transfers ($\mathcal{G}_t$). That is,  $\mathsf{WCC}(u_b,u_s, \mathcal{G}_t) \lor \mathsf{WCC}(u_b,\allowbreak u_s, \mathcal{G}_p))$ holds.

\noindent
\textbf{Rule 5.} We define \textit{shill score} as the ratio of the number of times $u_b$ participates in an auction created by $u_s$ and the total number of auctions that $u_b$ participated in.
In our detection approach, the \textit{shill score} must be greater than $\mu$, another empirically determined threshold.
We set $\mu=0.8$ for our analysis.

\vspace{0.5mm}
\noindent
$\blacktriangleright$ \textbf{Quantitative analysis.}
We detected \num{\shillbiddetected} instances of shill bidding across \num{\collectionsshillbidded} collections involving \num{\usersshillbidded} users in all \nftm s except \axie and \cryptopunks.
We estimate \textit{shill\_profit} as the profit made by the seller due to shill bidding.
Specifically, assume legitimate bidders place bids on an item first, and then shill bidding drives the price up.
If $b_l$ is the offer made by the last legitimate bidder before shill bidding  starts, and the item is finally sold at $b_s$ due to artificial inflation, we compute $(b_s - b_l)$ as the \textit{shill\_profit}.
According to our analysis, malicious sellers have collected a cumulative profit of \$\num{\totalshillbidsellerprofit} USD from all the shill bidding instances detected.
In \Fig{fig:shill_bid}, we show the frequency of shill bidding instances discovered across collections where shill bidding has been detected.
The majority (\num{\shillbiddingcollectionone}) of the collections have just one instance of shill bidding, while almost all collections ({\num{\shillbiddingcollectionlessthantwenty}}) have fewer than $20$ shill bids.

The one exception is the official collection of \foundation, which seems to be heavily affected by shill bidding.
With \num{\shillbiddingfoundation} instances ($\FPeval{\v}{round(\shillbiddingfoundation/\shillbiddetected*100, 2)}\v\%$ of all instances detected) of shill bids in that collection alone, it becomes the one with the most number of shill bids on any individual collection.
Our model also reports frequent shill bidding activity on the official collection of \superrare (\num{\shillbiddingsuperrare} instances) and \cryptovoxels (\num{\shillbiddingcryptovoxels} instances), which is an \opensea verified collection with \numwithsuffix{\cryptovoxelsitems} items and a cumulative trading volume of \numwithsuffix{\cryptovoxelsvolume} ETH.

\vspace{0.5mm}
\noindent
$\blacktriangleright$ \textbf{Manual analysis.}
Since shill bidding often closely resembles legitimate bidding behavior, it is harder to detect than other malpractices.
Therefore, to remain conservative during ground-truth determination, we looked for the following conditions:
\textbf{(i)} the shill bidder $S$ placed at least $3$ bids in an auction, and
\textbf{(ii)} if the average price of the items that $S$ bought is $p$, and the average bid that $S$ placed in that auction is $b$, then $p \ll b$, and
\textbf{(iii)} $S$ never bought any NFT from that seller.
We manually verified  \num{\shillbiddingmanuallyanalyzed} reports that we randomly selected from the instances that our approach detected.
Out of these \num{\shillbiddingmanuallyanalyzed} cases, $\shillbiddingtruepositive$ show strong indications of being instances of shill bidding.
For the remaining $\FPeval{\v}{round(\shillbiddingmanuallyanalyzed-\shillbiddingtruepositive,  0)}\v$, we could not draw any definitive conclusion from the trading patterns alone.
We observed an interesting shill bidding case in \foundation, where the initial reserve price of an NFT was $2$ ETH.
The item was targeted by a shill bidder who bid $[3.3, 4.4, 5.5, 6.71, 8.14]$ ETH on that item, thereby making the item finally sell at $9$ ETH.
However, all the NFTs owned by the bidder were worth between $(0, 2]$ ETH, and the bidder never bought any items from that seller.

\subsubsection{Bid Shielding}
\label{sec:bid_shielding}

\begin{figure}
	\centering
	\centering
	\includegraphics[width=0.6\linewidth]{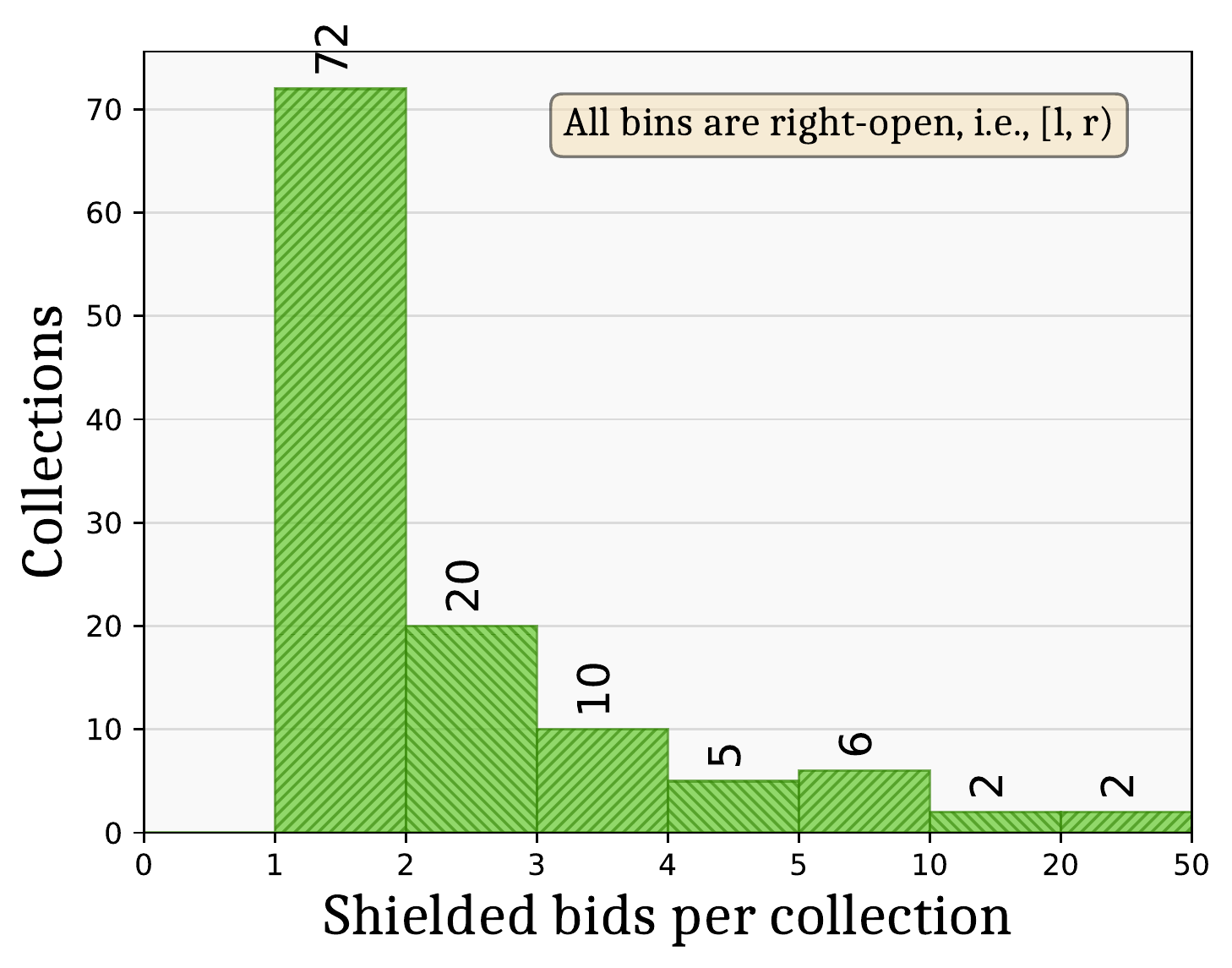}
	\vspace{-5mm}
	\caption{\small Distribution of bid shielding across collections}
	\label{fig:bid_shield}
	\vspace{-7mm}
\end{figure}

In bid shielding, a malicious bidder $u_2$ guards a low bid, possibly from a colluding bidder $u_1$, with a bid high enough to deter legitimate bidders from placing any additional bids.
Immediately before the auction ends, $u_2$ retracts the bid, thus uncovering the low bid from $u_1$ to let her win the auction.

\vspace{0.5mm}
\noindent
\textbf{Detection.} We apply the following heuristics to detect  instances of bid shielding in NFTMs.
If for two users $u_1$ and $u_2$, $\mathsf{bid}(u_1, p_1, t_1, id, a)$, $\mathsf{bid}(u_2, p_2, t_2, id, a)$ and  $\mathsf{cancel\_bid}(u_2, p_2, t_3, id, a)$ hold, then $u_2$ is shielding a bid from $u_1$ if:

\noindent
\textbf{Rule 1.} For all bids $\{\mathsf{bid}(u_i, \_, t_i, id, a)\}_{i=1}^{n}$ placed on asset $a$, $t_3$ > $t_i$ holds, \ie, no new bid was placed after $u_2$ retracted her bid on asset $a$.

\noindent
\textbf{Rule 2.} $u_1$ won the auction with id $id$. That is,  $\mathsf{win}(u_1,p_1,t_4,id, a)$ holds, and $u_1 \neq u_2 \land p_1 < p_2$.

\vspace{0.5mm}
\noindent
$\blacktriangleright$ \textbf{Quantitative analysis.}
We detected a total \num{\bidshieldsdetected} instances of bid shielding across \num{\collectionsbidshielded} collections involving \num{\usersbidshielded} users only in \opensea.
It is expected, because other \nftm s implement bidding policies (\Sec{sec:ecosystem}) to deter such malpractices, for example, on-chain bids, removal of the the previous bid when outbid, \etc{}
We compute \textit{shielded\_bid\_difference}, the difference in the bid amounts of the two colluding parties, the potential bid shielder and the auction winner.
While the minimum \textit{shielded\_bid\_difference} amount was \$\num{\bidshieldingminamount} USD, the maximum was as high as \$\num{\bidshieldingmaxamount} USD for one of the token in \mirandusvaults, a verified collection.
Additionally, all \num{\bidshieldsdetected} instances together shielded a total of \$\num{\totalshieldedbidamount} USD worth of bids.
\Fig{fig:bid_shield} shows the number of instances of bid shielding discovered across collections where bid shielding has been detected.
For most of the collections (\num{\bidshieldingcollectionlessthanten} out of \num{\collectionsbidshielded}), we find less than ten instances of bid shielding per collection.
\ens (ENS), a popular \ethereum name lookup service, makes it to the top of the list with \num{\bidshieldcountens} bid shielding instances.
Another notable finding in this category was the \cryptovoxels collection.
We noticed several complaints by \cryptovoxels collectors on their \discord server about the recent increase of bid shielding activity.
According to our analysis, \$\num{\bidshieldamountcryptovoxels} USD worth of bids were shielded by \num{\bidshieldcountcryptovoxels} instances of bid shielding, which corroborates this prior observation.
Our results show that bid shielding is frequent in verified collections as $\FPeval{\v}{round(\bidshieldinginverifiedcollections/\collectionsbidshielded*100, 2)}\v\%$ (\num{\bidshieldinginverifiedcollections} out of \num{\collectionsbidshielded}) of the bid shielded collections were verified. 

\vspace{0.5mm}
\noindent
$\blacktriangleright$ \textbf{Manual analysis.}
We have manually verified randomly chosen \num{\bidshieldingmanuallyanalyzed}  instances flagged by our analysis.
During manual analysis, we mark an instance as a \textit{true positive} if 
\textbf{(i)} the potential bid shielder $B$ and the (colluding) auction winner $W$ are the last two highest bidders on that auction in that order, and 
\textbf{(ii)} $B$ cancels her bid just before ($\leq 2$h) the auction ends, and 
\textbf{(iii)} during the auction, they never outbid each other.
Out of the \num{\bidshieldingmanuallyanalyzed} instances, our manual analysis confirms \num{\bidshieldingtruepositive} such instances as true positives.
Our detection model produced some false positives because it does not take into account the last condition listed above.
Let $b_i$ and $w_i$ be the bids from $B$ and $W$, respectively.
Now, first they outbid each other, \ie, $b_1 \rightarrow w_1 \rightarrow b_2 \rightarrow w_2 \rightarrow b_3$, and then $B$ removes $b_3$ at the last moment (possibly $B$ just changes her mind).
This is \textit{not} a bid shielding scenario, as the bids from $B$ drove up the price for $W$. This would not happen in a bid shielding scenario as $B$ and $W$ are colluding.
However, the first two conditions are still met, and therefore our model incorrectly flags this case. 
We also observed that most of bid shielding activities are performed in verified collections, such as \cryptovoxels, ENS, \etc, as they are popular and in high demand.

	\section{Related Work}
\label{sec:related_work}

To the best of our knowledge, we are the first to perform an in-depth study of security and privacy risks in the NFT ecosystem.
Our paper fits into the recent line of work on cryptoeconomic attacks %
in decentralized finance (\defii) systems.
The transparency of blockchains opens up the possibility of launching economic attacks by manipulating the market.
Since uncommitted \ethereum transactions and their gas bids are visible to other network participants, an attacker can offer a higher gas price to get their malicious transactions mined early in a block, before the victim transaction. 
This behavior is called \textit{front-running}~\cite{Eskandari20}.
The authors in~\flashboys~\cite{Daian19} demonstrated how arbitrage bots front-run transactions in decentralized exchanges (DEX) to generate non-trivial revenues.
\textit{Sandwich attacks} take this idea a step further by both front- and back-running victim transactions.
Zhou \etal~\cite{Zhou20} quantified the probability of being able to perform such an attack and the profits it can yield.
In fact, a recent paper~\cite{Qin21} reported the profit extracted from the blockchain to be a staggering $\$28.8$M USD in just two years, leveraging sandwiching, liquidation, and arbitrage.
The authors also measured the prevalence of other profit-making operations, \eg, \textit{clogging} and \textit{private mining}.
Another \defii{} trading instrument, \textit{flashloans}, allows a borrower immediate access to a large amount of funds without offering any collateral, under the condition that the loan needs to be repaid in the same transaction.
Qin \etal~\cite{Qin20} analyzed how flashloans have been used to execute arbitrage and oracle manipulation attacks, and they presented a constrained optimization framework to cleverly choose the attack parameters that maximize the profit. 
\defiposer~\cite{Zhou21} proposes trading algorithms to generate profit by crafting complex \defii{} transactions, both with and without flashloans.
Recent research~\cite{Kamps18,Gandal18,Xu2019} has also characterized and quantified \textit{pump-and-dump}, a price manipulation scheme that attempts to inflate the price of a crypto asset by spreading rumors and misinformation.

	\section{Conclusion}
\label{sec:conclusion}

This paper conducts the first systematic study of the emerging NFT ecosystem on \totalmarketplaces{} top NFT marketplaces (\nftm).
First, we perform a large-scale data collection from various sources, \viz, \ethereum mainnet,
\nftm{} websites, and their documentation.
We compile a comprehensive list of design weaknesses originating from the \nftm s and external entities, which often lead to financial consequences.
Further, we develop models to detect common trading malpractices, and quantify their prevalence in these marketplaces.

	\bibliographystyle{plain}
	\bibliography{bibs/conferences,bibs/nft,bibs/sailfish}
	\appendix
	\section{Charts}
\label{sec:charts}

\begin{figure}[h]
	\setlength{\fboxsep}{0pt}%
	\fbox{\includegraphics[width=0.8\linewidth]{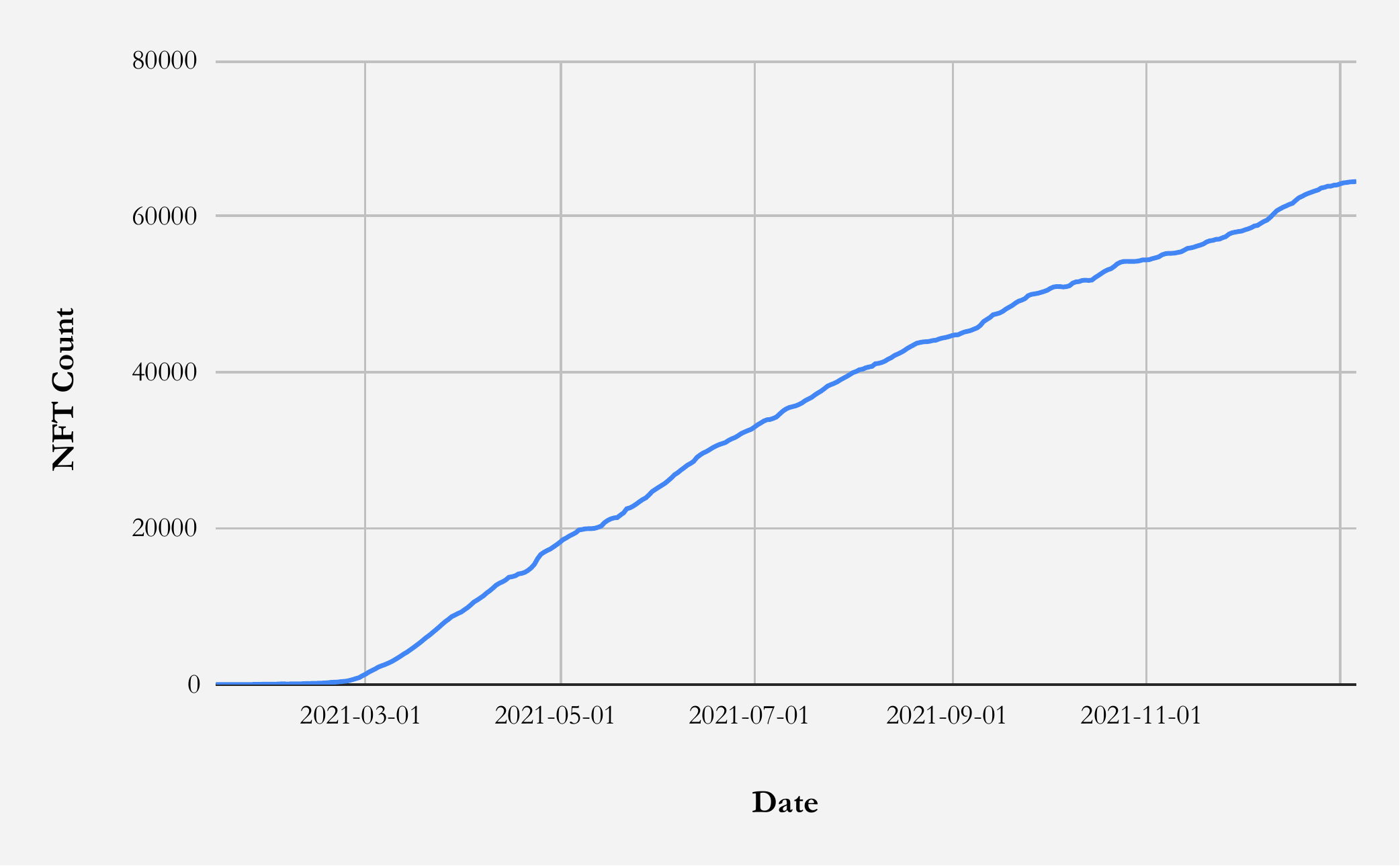}}
	\caption{Count of NFTs escrowed by \foundation over time}
	\label{fig:foundation_nft_count}
\end{figure}

\begin{figure}[h]
	\setlength{\fboxsep}{0pt}%
	\fbox{\includegraphics[width=0.8\linewidth]{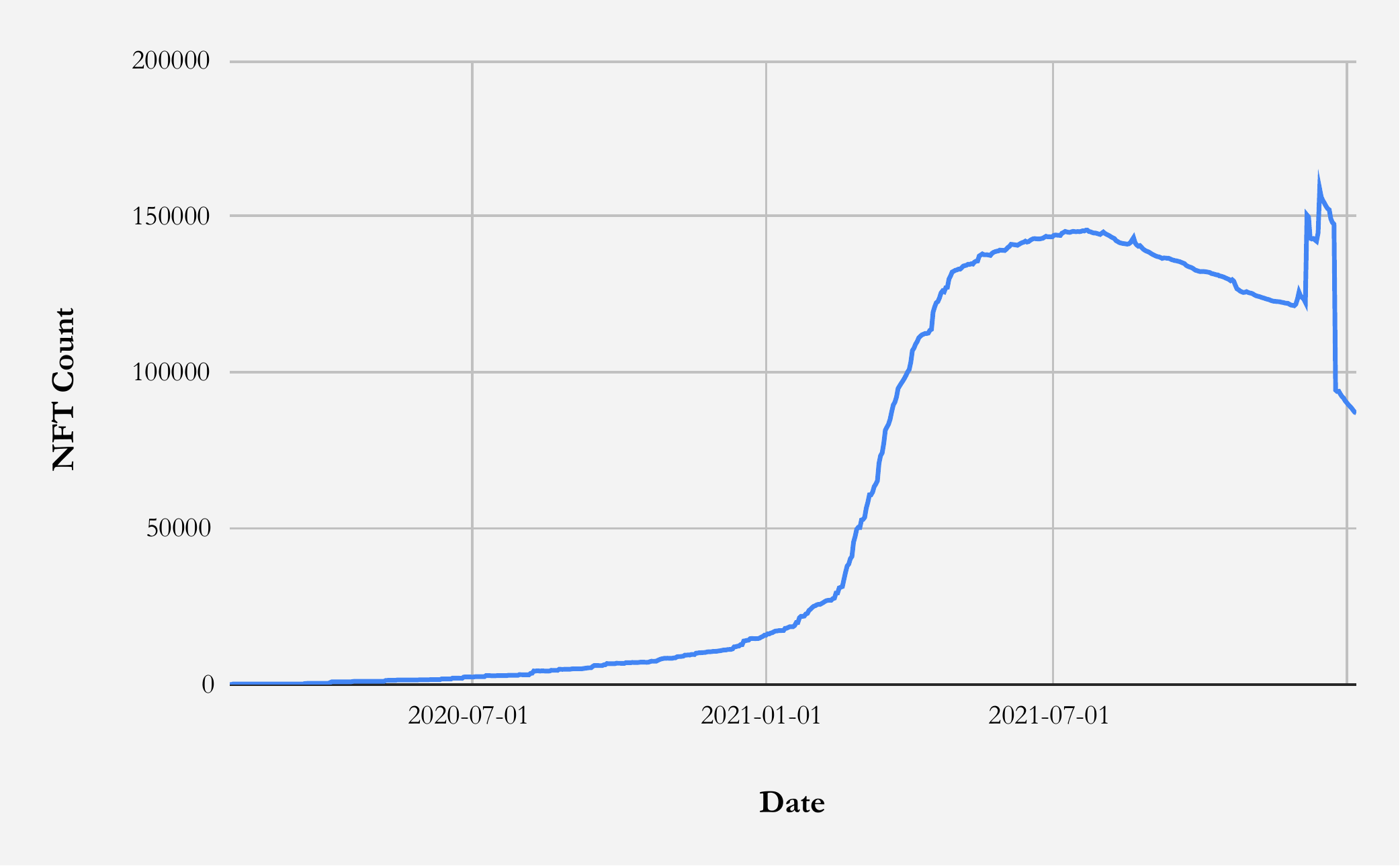}}
	\caption{Count of NFTs escrowed by \nifty over time}
	\label{fig:nifty_nft_count}
\end{figure}

	\section{Analysis of Top-15 NFT Sales}
\label{sec:top_sales}

In this section, we first discuss a few desirable properties of an NFT ecosystem.
Then, we analyze the top \topsales{} NFT sales by price with respect to those desirable properties.
We report a number of interesting observations with regards to the corresponding sales transactions.
Note that collecting information about our high-profile NFT sales was challenging since as is no source of ground truth. We primarily utilized four main sources: \textbf{(a)} search engines, \textbf{(b)} hashtag search in \twitter, \textbf{(c)} searches on \reddit, and \textbf{(d)} the \bc.

\begin{table}[ht]
	\centering
	\scriptsize
	\begin{tabular}
		{R{0.30\columnwidth}		%
	     C{0.10\columnwidth}		%
		 C{0.12\columnwidth}		%
		 C{0.10\columnwidth}		%
		 C{0.10\columnwidth}}   	%
	
		\toprule
		\textbf{NFT} & 
		\rotatebox{90}{\parbox{1.3cm}{\textbf{Sale Price}}} &
		\rotatebox{90}{\parbox{1cm}{\textbf{Sold on}}} &
		\rotatebox{90}{\parbox{1.3cm}{\textbf{Transferred\\on}}} &
		\rotatebox{90}{\parbox{1cm}{\textbf{Proof of\\Purchase}}} \\
		\midrule
		
		\rowcolor{black!10} 1. Beeple's Everydays & $69.30$M & 03/11/21 & 03/13/21 & \cross \\
		
		2. CryptoPunk $\#7523$ & $11.80$M & 06/10/21 & 07/15/21 & \cross \\
		
		\rowcolor{black!10} 3. CryptoPunk $\#7804$ & $7.56$M & 03/11/21 & 03/11/21 & \tick \\
		
		4. CryptoPunk $\#3100$ & $7.51$M & 03/11/21 & 03/11/21 & \tick \\
		
		\rowcolor{black!10} 5. Beeple's Crossroad & $6.66$M & 02/24/21 & N/A & \cross \\
		
		6. Beeple's OceanFront & $6.00$M & 03/23/21 & N/A & \cross \\
		
		\rowcolor{black!10} 7. CryptoPunk $\#5217$ & $5.44$M & 07/30/21 & 07/30/21 & \tick \\
		
		8. WWW source code & $5.43$M & -- & 09/10/21 & \cross \\
		
		\rowcolor{black!10} 9. CryptoPunk $\#7252$ & $5.30$M & 08/24/21 & 08/24/21 & \tick \\
		
		10. Snowden's StayFree & $5.27$M & 04/16/21 & 04/16/21 & \tick \\
		
		\rowcolor{black!10} 11. Save 1000s of Lives & $5.10$M & 05/08/21 & 05/08/21 & \tick \\
		
		12. CryptoPunk $\#2338$ & $4.40$M & 08/06/21 & 08/06/21 & \tick \\
		
		\rowcolor{black!10} 13. Micah's Replicator & $4.10$M & -- & 04/27/21 & \cross \\
		
		14. Fidenza $\#313$ & $3.30$M & 08/23/21 & 08/23/21 & \tick \\
		
		\rowcolor{black!10} 15. Jack Dorsey's Tweet & $2.90$M & 03/22/21 & N/A & \cross \\
		
		\bottomrule
	\end{tabular}
	\caption{Top $15$ most expensive NFT sales in descending order of sales price.
	Dates are in MM/DD/YY format.
	Missing information is denoted by `--', and N/A denotes that the corresponding event has not taken place yet.}
	\label{tbl:top_sales}
	\vspace{-3mm}
\end{table}

\subsection{Desirable properties of the ecosystem.}
Since NFTs are built around cryptocurrency and \bc, it is not unfair to expect the ecosystem to draw on the benefits offered by those technologies.
We identify the following properties that an \nftm{} protocol must hold in order to set it apart from traditional e-commerce platforms like Amazon, eBay, \etc{}
In other words, an \nftm{} that lacks in one or more of the following benefits should be deemed less valuable as a new platform---\textbf{(P1)} \textbf{Decentralization.}
NFTs should be stored on \bc{} to ensure persistence, censorship-resistance, immutability, and public verifiability of the asset-ownership record.
Keeping verifiability in mind, \ercnonfungible{} standard offers the \texttt{ownerOf()} API which returns the address of the current owner given a \texttt{\_tokenId}.
Since transfer events alter the ownership of an NFT, all those events should be recorded on-chain for an NFT to be verifiable.
Blockchain transaction history allows one to track when the NFT was created, who the previous owners were, and how much it was traded for each time.
If an \nftm{} protocol escrows a token to a wallet $W$, no sale record is emitted on the \bc, except the one where the current owner withdraws the token from $W$ to her own wallet.
Such a model makes the token opaque and unverifiable.
Also, as opposed to \bc, if the ownership record is stored in a centralized database, it is susceptible to tampering, censorship, and prone to disappear if the database goes away. 
An \nftm{} protocol should not sacrifice any of these guarantees in its design.
\textbf{(P2)} \textbf{Crypto payment.}
The payment toward the NFT trade must be made in cryptocurrency, \eg, a primary token like \ether{} (ETH), or a secondary token like Wrapped \ether{} (WETH), \etc{}
\textbf{(P3)} \textbf{Trustless trading.}
The trade must happen in a trustless manner, without relying on a third-party $T$ (other than the buyer $B$ and the seller $S$) who mediates either the transfer of assets or the payment.
For example, a protocol where $T$ escrows the token from $S$, accepts the payment from $B$, and then exchanges the token and the payment---beats the very purpose of decentralization by making the parties put trust on $T$.
\textbf{(P4)} \textbf{Atomic swap.}
The transfer of assets and the Crypto payment must take place in the same transaction in an \textit{atomic} manner, \ie, either both succeed, or both fail.
Atomicity enables two mutually distrusting parties to get involved in a trade without risking losing the asset or the funds.
\textbf{(P5)} \textbf{Token minting.}
A token has to be minted before or at the time of sales (\eg, lazy-minting), but not after.
This must hold because, in a trustless setting, an NFT cannot technically be sold, or transferred, unless it exists when the trade is executed.

\subsection{Analysis of top sales}
In \Tbl{tbl:top_sales}, we consolidate the sales information available in the public domain. As can be seen, at least half of the sales in \Tbl{tbl:top_sales} violate one or more of our desirable properties. We indicate the principle(s) a specific sale violates inside the parentheses. 

For Beeples's Everydays, the highest valued ($\$69.3$M) sale, we found the ownership transfer record one the \bc. However, payment is not present (\textbf{P4}). In other words, we failed to find any evidence  that $\$69.3$M was indeed transferred from MetaKovan (the buyer) to Beeple (the artist and seller).

For the second-largest transaction, several tweets~\cite{cryptopunk-7523-sale1,cryptopunk-7523-sale2} indicated that CryptoPunk $\#7523$ was sold on June $10, 2021$. However, the actual transfer on the \bc{} took place more than a month later (July $15, 2021$) (\textbf{P4}). Moreover, the transfer transaction had a value of zero ETH, thus making it impossible to confirm the reported sales amount ($\$11.8$M).

Beeple's CrossRoad, which was reportedly sold for $\$6.6$M, was traded on \nifty, which uses an escrow contract (\textbf{P3}). Thus, it does not have any sale or transfer record on the \bc{} (\textbf{P1}). Likewise, no public history exists for the sale of Beeple's OceanFront (\textbf{P1}, \textbf{P3}). In fact, querying the \bc{} with \texttt{ownerOf(\_tokenId)} returns the address of the \nifty gateway as its current owner.

For both the WWW source code and Micah's Replicator, the advertised payment amounts are not verifiable from the \bc{} transfer records (\textbf{P4}). And \twitter CEO Jack Dorsey's first tweet was sold on a platform called \valuables. This marketplace also uses an escrow contract, and, therefore, no sales or transfer details are publicly available (\textbf{P1}, \textbf{P3}). At the time of writing, querying the \valuables token contract on the \polygon sidechain returns an address owned by the platform as the owner.

	\section{Non-Technical Aspects of the Ecosystem}
\label{sec:non_technical_aspects}

In this section, we discuss a few non-technical questions surrounding the NFT ecosystem.
While we don't claim to be legal experts nor do we offer any tax advise, we frequently encountered certain issues during our work and wanted to bring them to our readers' attention.

\vspace{0.5mm}
\noindent
\textbf{Misconceptions around NFT purchase.}
Here we clarify some frequent misperceptions of the users interacting with the NFT space. 
\textbf{(i) Originality.}
Since digital artworks are infinitely and identically reproducible, the ``originality'' of a piece of art in the NFT world is ascertained by the \smart{} managing the corresponding token.
Unfortunately, fraudsters have been able to trick victims into buying NFTs pointing to someone else's art, for example, by deploying their own \smart s.
The NFT infrastructure is unable to provide any technical solution to this issue.
\textbf{(ii) Ownership.}
NFTs are used to introduce the concept of ``ownership'' to digital art. However, 
what that form of ownership actually means is somewhat subjective to individual's interpretations.
When an NFT is purchased, what the buyer really purchases is the NFT ``token'' on a \bc.
Whether and how that purchase translates to the ownership of the linked digital asset is debatable.
\textbf{(iii) Copyright.}
``Copyright'' grants the owner of the copyright the rights to control (a) the manufacture of copies of the original piece, (b) the sale, licensing, or transferring of the copyright itself, and (c) who can produce ``derivatives.''
Merely purchasing an NFT does not transfer the copyright to the buyer.
In one common scenario,  the buyer posts the linked artwork to social media. This essentially creates a digital copy of the art. Therefore, this might  infringe on the copyright of the artist, unless the terms of sale (ToS) explicitly allows the NFT buyer to do so.
\textbf{(iv) Terms of sale.}
A frequent misconception in the NFT space is that the ToS are encoded in the \smart.
Smart contracts are executable code.
Therefore, they can enforce certain aspects related to a trade, such as the sales price and royalty. 
But what if the seller were to include a term that precludes buyers from using the underlying digital art for commercial purposes? A \smart{} cannot enforce that provision.
A seller would have to resort to traditional methods of enforcement, \eg, demand letters, litigation, \etc

\vspace{0.5mm}
\noindent
\textbf{Valuation of NFT collections.}
The value people place in a work of art is largely subjective.
In the past, the price of an artwork has typically been decided by community consensus.
Moreover, a (valuable) artwork typically has a rich history associated with it. 
The fact that the NFT market is so young means that such history is not available. As a result, a certain amount of hype (and maybe market maniulation) drives up prices.
For example, the descriptions of a large number of NFT projects are rife with hyperbole, and it is not uncommon to find token owners on \twitter and \reddit providing long explanations of why a token they own is particularly meaningful.
Since NFTs lack any intrinsic value, it is non-trivial for a newcomer to judge its true merit.
Hence, they can easily fall prey of such promotions, and sometimes run into exit scams.

\vspace{0.5mm}
\noindent
\textbf{Tax implications.}
Law practitioners seem to agree that the purchase or sale of NFTs is a taxable transfer of property, and is therefore subject to capital gains tax.
We noticed users gifting NFTs to others, which might trigger a gift tax.
Unfortunately, taxation on NFTs is still a gray zone, as the tax laws are unclear and classic securities laws need to be reapplied.
\nftm s are open markets, and cross-border sales can make matters complicated, as NFT buyers and sellers  have to deal with different jurisdictions' tax regimes.
Also, NFT trades could violate U.S. sanctions law, which prevents U.S. residents or citizens from conducting business with individuals or entities from sanctioned nations.
Experts are not even ruling out the possibility of NFTs being used for money laundering to support illicit activities.
While taxation regulations are already complicated, none of our examined marketplaces help user to remain tax-compliant by generating tax forms, \eg, 1099K.
Instead, we see disclaimers such as \opensea's terms of service (ToS), which states: ``\textit{You are solely responsible for determining what, if any, taxes apply to your Crypto Assets transactions.
	Neither OpenSea nor any other OpenSea Party is responsible for determining the taxes that apply to Crypto Assets transactions}.''
Given the lack of clarity and support, it is  possible for unsuspecting, law-abiding users to inadvertently violate tax rules while interacting with these marketplaces.

\noindent
\textbf{Lack of support for selling physical assets.}
Though NFTs are being used to trade physical assets in limited cases, the current state of the affairs not only violates the basic principles of \bc{} sales, but also it gives rise to the potential of an abuse.
\nftm s enable two mutually distrusting parties to execute trades in a trustless environment while retaining their anonymity.
However, delivery of physical goods requires sharing the details of the buyer with the seller.
In a centralized marketplace, \eg, Amazon, the platform itself acts as the trusted third-party (TTP) that protects the buyer information from the seller, often handling the delivery on the seller's behalf.
Unfortunately, no current \nftm{} offers such a service.
In fact, even if they would do so, that would violate the spirit of a trustless, peer-to-peer marketplace.
The alternative, which is the current practice, is to have the buyer share her contact details directly with the seller.
For example, it is not uncommon to find NFTs sold by photographers where they promise the buyer a physical print of the photo, and, therefore, they request the buyer to email their address to the seller.
Needless to say, this poses a significant threat to the buyer's privacy.
In addition, the absence of a TTP makes arbitration harder in case of any disputes, \eg, non-delivery of the purchased asset.
To summarize, 
no current marketplace protocol satisfies all three desirable yet mutually conflicting requirements, \viz, trustlessness, decentralization, and anonymity. Thus, NFT markets are not entirely suitable as platforms to sell physical assets.

	\section{Contract Addresses}
\label{sec:contract_addresses}
We provide the \ethereum addresses of the important contracts used in this paper in \Tbl{tbl:contract_addresses}.

\begin{table}[h]
	\footnotesize
	\begin{tabular}{llc}
		\hline
		\multicolumn{1}{|l|}{\thead{Marketplace}} & \multicolumn{1}{l|}{\thead{Purpose}} & \multicolumn{1}{l|}{\thead{Contract Address}} \\ \hline
		\multicolumn{1}{|l|}{\opensea} & 	\multicolumn{1}{l|}{Marketplace} & \multicolumn{1}{l|}{ 0x7be8076f4ea4a4ad08075c2508e481d6c946d12b}                 \\ \hline
		\multicolumn{1}{|l|}{\opensea}            & \multicolumn{1}{l|}{Token}        & \multicolumn{1}{l|}{0x495f947276749ce646f68ac8c248420045cb7b5e}                 \\ \hline
		\multicolumn{1}{|l|}{\axie}            & \multicolumn{1}{l|}{Marketplace}        & \multicolumn{1}{l|}{0xf4985070ce32b6b1994329df787d1acc9a2dd9e2}                 \\ \hline
		\multicolumn{1}{|l|}{\axie}            & \multicolumn{1}{l|}{Token}        & \multicolumn{1}{l|}{0xf5b0a3efb8e8e4c201e2a935f110eaaf3ffecb8d}                 \\ \hline
		\multicolumn{1}{|l|}{\cryptopunks}            & \multicolumn{1}{l|}{Marketplace}        & \multicolumn{1}{l|}{0xb47e3cd837ddf8e4c57f05d70ab865de6e193bbb}                 \\ \hline
		\multicolumn{1}{|l|}{\cryptopunks}            & \multicolumn{1}{l|}{Token}        & \multicolumn{1}{l|}{0xb47e3cd837ddf8e4c57f05d70ab865de6e193bbb}                 \\ \hline
		\multicolumn{1}{|l|}{\rarible}            & \multicolumn{1}{l|}{Marketplace}        & \multicolumn{1}{l|}{0x9757f2d2b135150bbeb65308d4a91804107cd8d6}                 \\ \hline
		\multicolumn{1}{|l|}{\rarible}            & \multicolumn{1}{l|}{Token}        & \multicolumn{1}{m{4.83cm} |}{\shortstack[l]{0x60f80121c31a0d46b5279700f9df786054aa5ee5 \\ 0xd07dc4262bcdbf85190c01c996b4c06a461d2430 \\ 0x6a5ff3ceecae9ceb96e6ac6c76b82af8b39f0eb3}}                 \\ \hline
		\multicolumn{1}{|l|}{\superrare}            & \multicolumn{1}{l|}{Marketplace}        & \multicolumn{1}{m{4.83cm}|}{\shortstack[l]{0x2947f98c42597966a0ec25e92843c09ac17fbaa7 \\ 0x8c9f364bf7a56ed058fc63ef81c6cf09c833e656 \\ 0x65b49f7aee40347f5a90b714be4ef086f3fe5e2c }}                 \\ \hline
		\multicolumn{1}{|l|}{\superrare}            & \multicolumn{1}{l|}{Token}        & \multicolumn{1}{l|}{0xb932a70a57673d89f4acffbe830e8ed7f75fb9e0}                 \\ \hline
		\multicolumn{1}{|l|}{\sorare}            & \multicolumn{1}{l|}{Marketplace}        & \multicolumn{1}{l|}{0xaeb960ed44c8a4ce848c50ef451f472a503456b2}                 \\ \hline
		\multicolumn{1}{|l|}{\sorare}            & \multicolumn{1}{l|}{Token}        & \multicolumn{1}{m{4.83cm}|}{\shortstack[l]{0x629a673a8242c2ac4b7b8c5d8735fbeac21a6205 \\ 0x9844956f1d45996aa8d322f3483cc58abe34d449 \\ 0xd2c98d651a02e34c279ed470a1447a36aa0423ee}}                 \\ \hline
		\multicolumn{1}{|l|}{\foundation}            & \multicolumn{1}{l|}{Marketplace}        & \multicolumn{1}{l|}{0xcda72070e455bb31c7690a170224ce43623d0b6f}                 \\ \hline
		\multicolumn{1}{|l|}{\foundation}            & \multicolumn{1}{l|}{Token}        & \multicolumn{1}{l|}{0x3b3ee1931dc30c1957379fac9aba94d1c48a5405}                 \\ \hline
		\multicolumn{1}{|l|}{\nifty}            & \multicolumn{1}{l|}{Marketplace}        & \multicolumn{1}{l|}{off-chain}                 \\ \hline
		\multicolumn{1}{|l|}{-}            & \multicolumn{1}{l|}{CelebrityBreeder}        & \multicolumn{1}{l|}{0xa33ab4b0c9905ebc4e0df5eb2f915bee728b8253}                 \\ \hline
		\multicolumn{1}{|l|}{-}            
		& \multicolumn{1}{l|}{Mirandus Vaults} 
		& \multicolumn{1}{l|}{0x495f947276749ce646f68ac8c248420045cb7b5e}
		\\ \hline
		\multicolumn{1}{|l|}{-}            
		& \multicolumn{1}{l|}{Gods Unchained} 
		& \multicolumn{1}{l|}{0x0e3a2a1f2146d86a604adc220b4967a898d7fe07}
		\\ \hline
	\end{tabular}
	\vspace{1mm}
	\caption{\ethereum addresses of important contracts.}
	\label{tbl:contract_addresses}
\end{table}

	\section{Fraudulent User Behaviors - Extended}
\label{sec:user_issues_extended}

\noindent
\textbf{Digital scarcity.}
Digital scarcity~\cite{digital-scarcity} is the limitation, typically imposed through software, to control the abundance of a digital resource.
The more abundant an asset is, the lesser becomes its intrinsic value.
Since NFTs are created by \smart s, it is possible to impose appropriate limitations to ensure scarcity, provided: \textbf{(i)} the rarity parameter is stored on-chain, and \textbf{(ii)} the contract uses the parameter to prohibit minting beyond promised limits.

Currently, most of the items that claim to be a \textit{limited edition}, or \textit{rare}---it is word of mouth, than any contract-level guarantee.
In fact, we found 
users complaining about a `supposed' limited edition item being minted beyond the promised limit.
\cryptomotors, a verified collection in \opensea, claims to have only $150$ GEN1 cars in circulation.
However, the rarity parameter (GEN) is stored off-chain inside the JSON metadata, making it impossible to enforce rarity at the contract level.
Additionally, the \texttt{totalSupply} parameter, which controls the total supply of a token, is also not fixed, which makes it possible to mint unlimited cars.

\vspace{0.5mm}
\noindent
\textbf{Giveaway scams.}
NFT giveaways are campaigns to distribute free NFTs in exchange for having users promote the newly launched collection on social media.
In giveaways scams, scammers lure the users of free NFTs, but ask for `small' fees to cover the gas cost.
In reality, the fee they ask for is several times greater than the gas cost required for the transfer.
Sometimes, NFT platforms use a fungible token as the native currency for their services.
For example, \nftartfinance~\cite{nft-art-finance} is powered by their platform token called \$NFTART.
In some scams, the scammers pretend to put either the NFT, or the platform token `on-sale'.
Users who fall for this send funds to the designated accounts, but never receive the NFT or the tokens in return.
Interestingly, there have also been instances where legitimate giveaways were targeted by scammers where they impersonated the `winner' by faking social media accounts, and had the reward transferred to their wallet, thus forfeiting the real winner.

\vspace{0.5mm}
\noindent
\textbf{Front-running.}
In a front-running attack, an attacker gets a malicious transaction mined before a victim by paying a higher gas price.
When a transaction is broadcast in the \ethereum network, it appears in the \textit{mempool}.
A replay attack synthesizes a malicious transaction from a profit-making mempool transaction, oftentimes just by copying the arguments verbatim---only to front-run the victim to bag the profit.
In reality, automated bots sniff the mempool for such profitable victims.
Since NFTs are managed by \smart s, those are susceptible to front-running.

Also, it has been shown that by merely front-running the \texttt{giveBirth} call~\cite{cryptokitties-frontrun} of the \cryptokitties token, an attacker would make a profit of $\$111$K USD.
In February $2021$, an attacker exploited a weakness in \cryptopunks's bid acceptance mechanism~\cite{cryptopunks-frontrun}, for which a bid that was supposed to be closed for $26.25$ ETH, returned only $1$ Wei ($=10^{-18}$ ETH) in profit due to being front-run.

\vspace{0.5mm}
\noindent
\textbf{Insider trading.}
An \textit{insider} is one who has access to some confidential information about publicly traded security.
Any trade involving an insider is an insider trade.
However, it is illegal when the investor leverages that information in deciding when to buy or sell the security, because it gives them an unfair advantage to make a profit from that information.
The regulatory gap in the NFT ecosystem surfaced out recently once again when an \opensea employee was found to be involved in an illegal insider trade in September $2021$.
Leveraging internal information, that employee bought NFT just before it was featured on the front page of the marketplace, and then sold it right after it soared in price---making a profit of $18.875$ ETH in total.

	\section{Artifacts}
\label{sec:artifacts}

The data related to the NFTs and NFT marketplace activities that we collected over a period of time for the purpose of this study is one of our claimed contribution.
Since the size of the data is in several terabytes, preparing it in a form usable by the reviewers, and then uploading it to an external service could take several hours to days.
We have already promised to release the code and the dataset upon acceptance.
Nevertheless, if the reviewers feel the need of having access to the dataset as part of the review process, we will make the requested subset available during the rebuttal phase.

	\section{Data Collection} 
In \Tbl{tbl:data_collection}, we provide the details of the types of data, \ie, assets and events we collected from different marketplaces and \bc{}.
\label{sec:data_collection}

\begin{table}[htbp]
	\begin{tabular}{p{0.2cm}|p{2.3cm}|p{4.5cm}}
		\toprule
		\rowcolor{black!10}\rotatebox{90}{\thead{Entity}}         & \thead{Attribute}                               & \thead{Description} \\ \midrule
			\multirow{8}{*}[-30mm]{\rotatebox{90}{\textbf{Asset}}} & Token contract address                         &  \ethereum address of the token contract that manages the asset           \\ \cmidrule{2-3} 
			& Token ID                                       & Integer ID that uniquely identifies the token among all the tokens managed by the token contract            \\ \cmidrule{2-3} 
			& Collection name                                &  Name of the collection that the NFT belongs to           \\ \cmidrule{2-3} 
			& Image URL                                      & URL of the resource (picture/video) that is pointed to by the NFT           \\ \cmidrule{2-3} 
			& Metadata URL                                   & URL of the metadata JSON if the token is \ercnonfungible-compliant and implements the metadata extension             \\ \cmidrule{2-3} 
			& Asset listing URL                              & URL of the listing page of that asset on the \nftm{} \dapp           \\ \cmidrule{2-3}
			& Source code availability of the token contract & Boolean flag indicating if the source code of the token contract is available in \etherscan            \\ \cmidrule{2-3} 
			& Verification status of the collection          &  Boolean flag indicating if the collection which this asset belong to is verified by the \nftm            \\ \midrule
			
			\multirow{7}{*}[-50mm]{\rotatebox{90}{\textbf{Event}}} &   Mint (Asset creation)                                         & Minter's (Creator's) address, minting time, asset being minted            \\ \cmidrule{2-3} 
			& Sell                                               & Seller's address, Buyer's address, Timestamp, Transaction hash, Asset being sold, Sell price (USD, ETH)            \\ \cmidrule{2-3}
			& Asset transfer & From address, To address, Asset being transferred, Timestamp, Transaction hash                                               \\ \cmidrule{2-3}
			& Auction start                                               & Asset on which the auction started, Auction creator, Timestamp, Transaction hash            \\ \cmidrule{2-3}            
			& Bid                                               & Bidder, Asset on which bid is placed, Auction creator, Bid amount (USD, ETH), Timestamp, Transaction hash            \\ \cmidrule{2-3}
			& Bid cancel                                               & Bidder, Asset on which bid is canceled, Auction creator, Cancel price, Timestamp, Transaction hash            \\ \cmidrule{2-3} 
			& Win                                                           & Winner, Asset being won, Auction creator, Sell price (USD, ETH), Timestamp, Transaction hash            \\ \cmidrule{2-3}
			& Auction end                                               & Asset for which the auction ended, Auction creator, Timestamp, Transaction hash            \\ \cmidrule{2-3} 
			& ETH transfer                                               & From address, To address, Amount (ETH), Timestamp, Transaction hash            \\ 
			\bottomrule
		\end{tabular}
		\vspace{1mm}
		\caption{Details of the data collected for this study.}
		\label{tbl:data_collection}
	\end{table}

\end{document}